\documentclass[useAMS,usenatbib]{mn2e}
\usepackage{graphicx}
\usepackage{booktabs}
\usepackage[usenames,dvipsnames,svgnames,table]{xcolor}
\usepackage{chngcntr}
\counterwithout{table}{section}
\usepackage{verbatim}
\usepackage{float}
\usepackage{pdflscape}
\usepackage{amsmath}
\begin{document}
\title[Radio galaxies in ZFOURGE/NMBS]{Radio galaxies in ZFOURGE/NMBS: no difference in the properties of massive galaxies with and without radio-AGN out to z=2.25}
\author[G.~Rees et al.]{G. A. Rees$^{1,2}$\thanks{This paper includes data gathered with the 6.5 meter Magellan Telescopes located at Las Campanas Observatory, Chile.}, L. R. Spitler$^{1,3}$, R. P. Norris$^{2}$, M. J. Cowley$^{1,3}$, C. Papovich$^{4,5}$, 
\newauthor K. Glazebrook$^{6}$, R. F. Quadri$^{4,5}$, C. M. S. Straatman$^{7}$, R. Allen$^{3,6}$, G. G. Kacprzak$^{6}$,
\newauthor I. Labbe$^{7}$, T. Nanayakkara$^{6}$,   A. R. Tomczak$^{4,5}$, K.-V. Tran$^{4,5}$.  \\
$^{1}$Macquarie University, Balaclava Road, Epping, NSW, Australia  \\
$^{2}$CSIRO Australia Telescope National Facility, PO Box 76, Epping, NSW, 1710, Australia \\
$^{3}$Australian Astronomical Observatories, PO Box 915 North Ryde NSW 1670, Australia \\
$^{4}$George P. and Cynthia Woods Mitchell Institute for Fundamental Physics and Astronomy, Texas A$\&$M University, College Station,\\   $~$TX, 77843-4242 USA\\
$^{5}$Department of Physics and Astronomy, Texas A$\&$M University, College Station, TX, 77843-4242 USA\\
$^{6}$Centre for Astrophysics \& Supercomputing, Swinburne University, Hawthorn, VIC 3122, Australia\\
$^{7}$Leiden Observatory, Leiden University, P.O. Box 9513, 2300 RA Leiden, The Netherlands}

\date{Accepted: XXXX. Received XXXX}
\pagerange{\pageref{firstpage}--\pageref{lastpage}} \pubyear{2014}

\maketitle
\label{firstpage}
\begin{abstract}

In order to reproduce the high-mass end of the galaxy mass-distribution, some process must be responsible for the suppression of star-formation in the most massive of galaxies. Commonly Active Galactic Nuclei (AGN) are invoked to fulfil this role, but the exact means by which they do so is still the topic of much debate, with studies finding evidence for both the suppression and enhancement of star-formation in AGN hosts. Using the ZFOURGE and NMBS galaxy surveys, we investigate the host galaxy properties of a mass-limited (M$_{\odot}$ $\ge$ 10$^{10.5}$ M$_{\odot}$), high-luminosity (L$_{1.4}$ $>$ 10$^{24}$ W~Hz$^{-1}$) sample of radio-loud Active Galactic Nuclei  to a redshift of z = 2.25. In contrast to low redshift studies, which associate radio-AGN activity with quiescent hosts, we find that the majority of z $>$ 1.5 radio-AGN are hosted by star-forming galaxies. Indeed, the stellar populations of radio-AGN are found to evolve with redshift in a manner that is consistent with the non-AGN mass-similar galaxy population. Interestingly, we find the radio-AGN fraction is constant across a redshift range of 0.25 $\le$ z $<$ 2.25, perhaps indicating that the radio-AGN duty cycle has little dependence on redshift or galaxy type. We do however see a strong relation between the radio-AGN fraction and stellar mass, with radio-AGN becoming rare below $\sim$ 10$^{10.5}$ M$_{\odot}$ or a halo-mass of 10$^{12}$ M$_{\odot}$. This halo-mass threshold is in good agreement with simulations that initiate radio-AGN feedback at this mass limit. Despite this we find that radio-AGN host star-formation rates are consistent with the non-AGN mass-similar galaxy sample, suggesting that while radio-AGN are in the right place to suppress star-formation in massive galaxies they are not necessarily responsible for doing so.

\end{abstract}
\begin{keywords}
galaxies: evolution, galaxies: high-redshift, galaxies: active, radio continuum: galaxies, infrared: galaxies, galaxies: stellar-content
\end{keywords}

\section{Introduction}

Super-massive black-hole (SMBH) accretion is now considered to be one of the primary regulators of galaxy evolution. Because of this, understanding the ways in which this process can begin, along with the impact it has on the surrounding environment is one of the key questions in extra-galactic astronomy. \\

It has been suggested that actively accreting SMBHs in the center of galaxies, now commonly referred to as a Active Galactic Nuclei (AGN), can be triggered in a variety of ways. Merger driven models \citep{Hopkins2008, RamosAlmeida2010, Karouzos2010, Sabater2013}, induce accretion by both disrupting existing cold gas reservoirs in the merging galaxies and via the direct inflow of gas from the merger event itself. Alternatively, there is evidence to support the triggering of AGN in non-merging galaxies (\citealt{Crenshaw2003}, \citealt{Schawinski2014a}, \citealt{Draper2012} and references therein) via supernova winds, stellar bars and the efficient inflow of cold gas from the intergalactic medium. \\

At the same time, models have explored the various forms of feedback from AGN: pressure, mechanical energy and heating \citep{McNamara2005, Nulsen2013}. However, the exact impact of these different processes is still a topic of much debate, with observations supporting both the the suppression \citep{Morganti2013, Nesvadba2010} and inducement \citep{Bicknell2000, Silk2010, Zinn2013, Karouzos2014} of star-formation by AGN activity. \\

Radio-AGN jets in particular are known to inject mechanical feedback that can heat gas and suppress star-formation. Given that radio-AGN tend to be found in massive galaxies \citep{Auriemma1977,Dressel1981}, it has been suggested that these processes may be responsible for the suppression of star-formation in high mass galaxies \citep{Springel2005,Croton2006}. Low redshift observations support this theory, finding that the majority of radio-AGN are located in high-mass quiescent galaxies. \citep{Best2005, Best2007, Kauffmann2008}. Higher redshift studies (1 $\le$ z $\le$ 2), which estimate redshift using the tight correlation observed between between K-band magnitude and redshift for radio-galaxies \citep{Laing1983} agree with this, finding that radio-AGN hosts are consistent with passively evolving stellar populations formed at z $>$ 2.5 \citep{Lilly1984, Jarvis2001}. \\

However, above z $\approx$ 2, radio-AGN hosts begin to show magnitudes greater than we might expect from passive evolution alone, indicating that at these redshifts we may at last be probing the formation period of massive radio-detected galaxies \citep{Eales1997}. There is also growing observational evidence that star-formation may be common in high redshift radio-AGN hosts \citep{Stevens2003,Rocca-Volmerange2013a}, and it has been suggested that we are witnessing these objects during the transition phase between the ignition of the radio-AGN and the quenching of star-formation by AGN feedback \citep{Seymour2012, Mao2014a}. Finally, observations at millimetre wavelengths support the existence large cold gas reservoirs in high-redshift, high-luminosity (L$_{1.4}$ $>$ 10$^{24}$ WHz$^{-1}$) radio-AGN hosts \citep{Emonts2014}, suggesting that at high redshifts the fuel for star-formation has not yet been removed by AGN feedback.\\

The suggestion that many high-redshift, high-luminosity radio-AGN are star-forming is interesting, as unlike their lower-luminosity Seyfert counterparts, these objects are rare at low redshifts (z $<$ 0.05)\citep{Ledlow2001,Keel2006,Mao2014b}. One simple explanation for the apparent increase in actively star-forming radio-AGN hosts with redshift, is that high-redshift quiescent galaxies are simply more difficult to detect. This is particularly true in optically selected radio-AGN samples, as at z $\ge$ 1.6 optical observations are probing the faint rest-frame UV part of the quiescent galaxy spectrum. Because of this, studies using samples of spectroscopically confirmed objects may be inherently biased towards star-forming objects. \\

This problem is much smaller for samples selected using detections in both radio and K-band, and the K-Z relation allows for reasonably accurate redshift measurements out to z$\sim$2.5. However, the relation is known to broaden significantly at high redshifts and lower radio fluxes \citep{Eales1997, DeBreuck2002}, making the study of radio-AGN host properties difficult in these regimes as the accuracy of the objects redshift becomes increasingly uncertain. \\

Finally, and perhaps most importantly, radio-galaxies identified using either optical spectroscopy or the K-Z relation typically lack a readily available control population free of radio-AGN to compare against. What is needed, is a rest-frame-optical (observed near-infrared) selected galaxy catalogue, supplemented with deep radio imaging and accurate redshift information in order to effectively study the effects of radio-AGN on their hosts at each epoc. \\

In this paper, we present an analysis of two deep near-infrared ({\it K}-band) surveys combined with high-sensitivity 1.4~GHz radio observations. By selecting radio-loud AGN (radio-AGN hereafter) in this way, we are able to produce a mass and radio-luminosity complete sample of galaxies across a photometric redshift range of 0.25 $<$ z $<$ 2.25, thereby minimizing our biases towards either star-forming or quiescent hosts. Furthermore, our observations include a large population of non-AGN galaxies at similar masses and redshifts to our radio-AGN sample, allowing us to directly compare star-formation rates between radio-AGN and non-AGN hosts in search of AGN feedback, while controlling for the effects of redshift and mass. \\

Section \ref{Data} of this paper describes our multi-wavelength data, as well as their unification into a single sample via cross-matching. Section \ref{Sec:UVJ} outlines our methods for classifying our sample into quiescent or star-forming hosts based on optical and Near-Infrared (NIR) stellar population emission and Section \ref{Sec:AGN} covers our process for determining whether or not a galaxy is hosting a radio-AGN. The final sections contain our analysis (Section \ref{Sec:Analysis}) and discussion (Section \ref{Sec:Discuss}) of the resultant sub-samples.

\section{Data}
\label{Data}
	\subsection{Near-Infrared observations}

\begin{figure*}
	\centering
	\includegraphics[width=2.0\columnwidth]{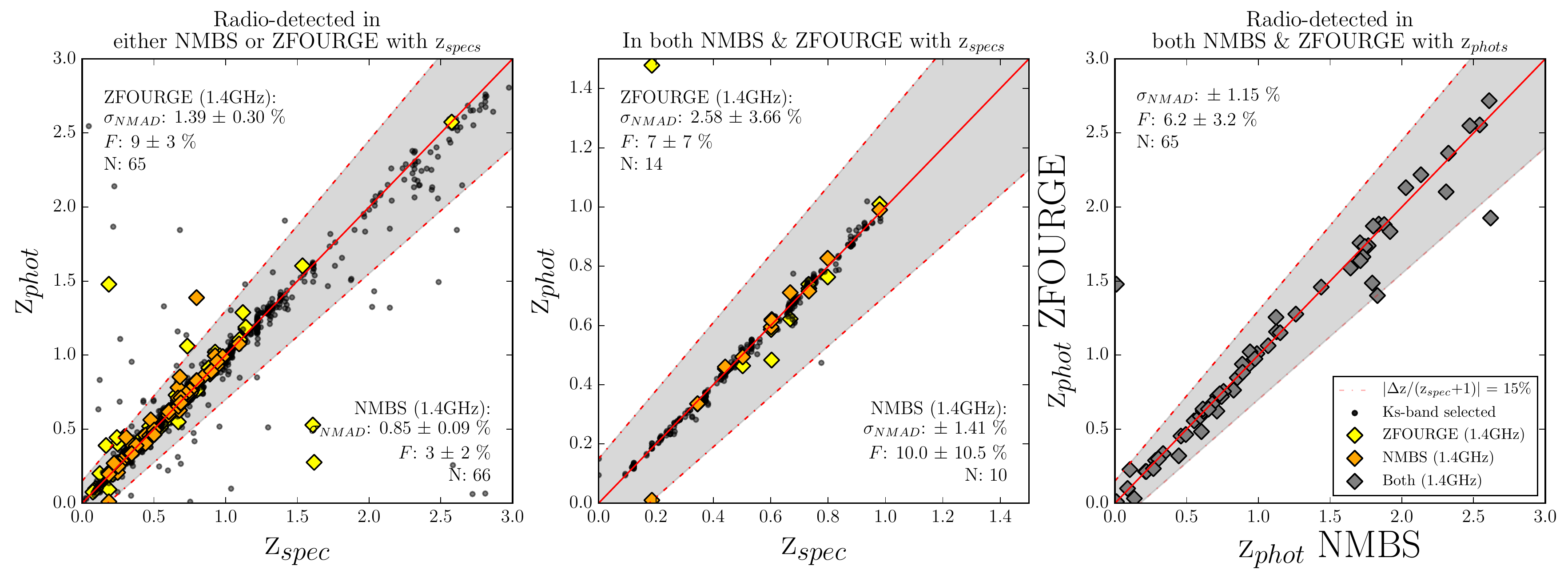} \\
	\caption{All: The Normalised Median Absolute Deviation of source redshifts ($\sigma_{NMAD}$), the catastrophic outlier fraction (F) where $\Delta$z/(z$_{spec}$+1) $<=$ 15\% and the number of points included in the sample (N). Left: A photometric versus spectroscopic redshift comparison for radio-detected objects in either ZFOURGE (yellow diamonds) or NMBS (orange diamonds). {\it Ks}-selected objects  with available spectroscopic redshifts from ZFOURGE are shown for reference (black circles). A 1:1 line is also plotted (solid red) along with the $\Delta$z/(z$_{spec}$+1) $<=$ 15\% limits for reference (grey area). Center: A photometric versus spectroscopic redshift comparison for radio-objects detected in both ZFOURGE and NMBS. Right: A photometric versus photometric redshift comparison for radio-objects detected in both ZFOURGE and NMBS (grey diamonds). While ZFOURGE in general suffers a slightly higher outlier fraction than NMBS in the leftmost panel this is largely due to a small population of poorly fitted faint sources that are undetected in NMBS. These sources are flagged and are not included in our analysis.}
	\label{fig:ZphotZspec}
\end{figure*}

\begin{figure}
	\centering
	\includegraphics[width=1.0\columnwidth]{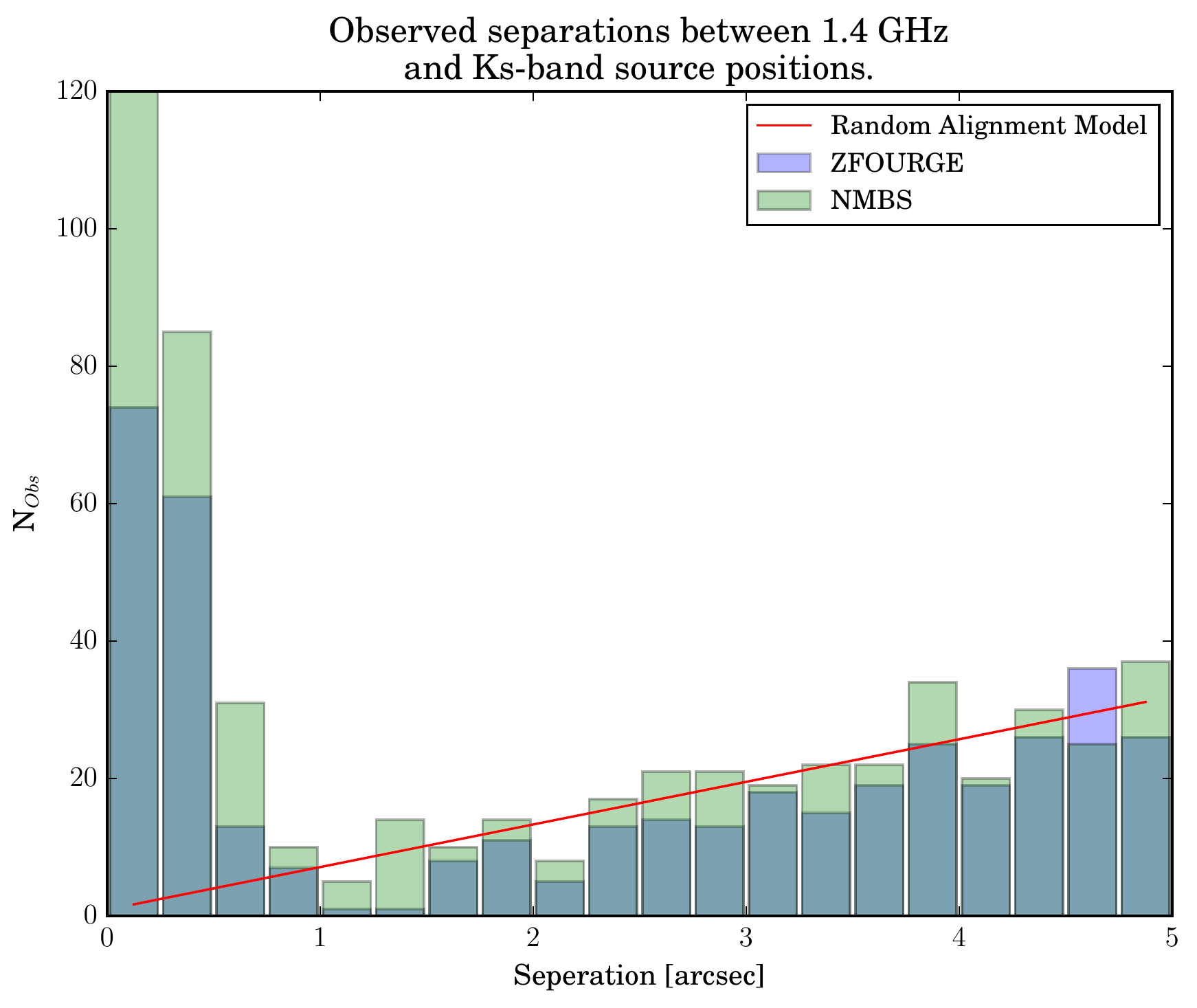} \\
	\includegraphics[width=1.0\columnwidth]{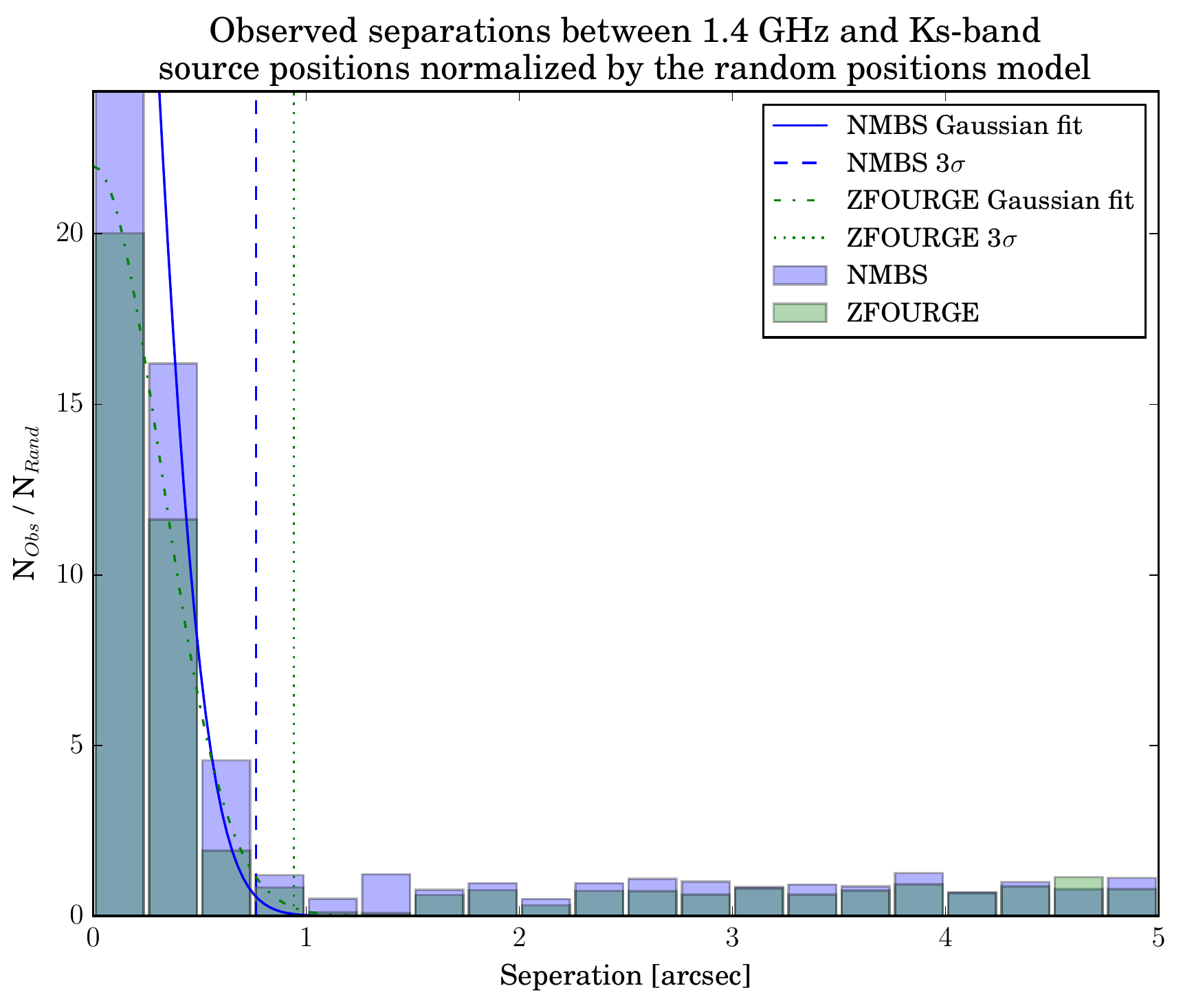} \\
	\caption{Top: Histogram detailing the number of radio to {\it Ks}-band pairs as a function of separation for the ZFOURGE (blue) and NMBS (green) surveys. The expectation for randomly positioned, point like sources shown (red line). Bottom: Same as above but normalised by the random expectation model. Gaussians are fitted to the residual ZFOURGE \& NMBS populations and their three sigma limits are shown (dotted and dashed vertical lines). We use a 1$\arcsec$ cross-matching radius for both surveys.}
	\label{fig:Seps}
\end{figure}

\begin{figure*}
	\centering
	\includegraphics[width=2.0\columnwidth]{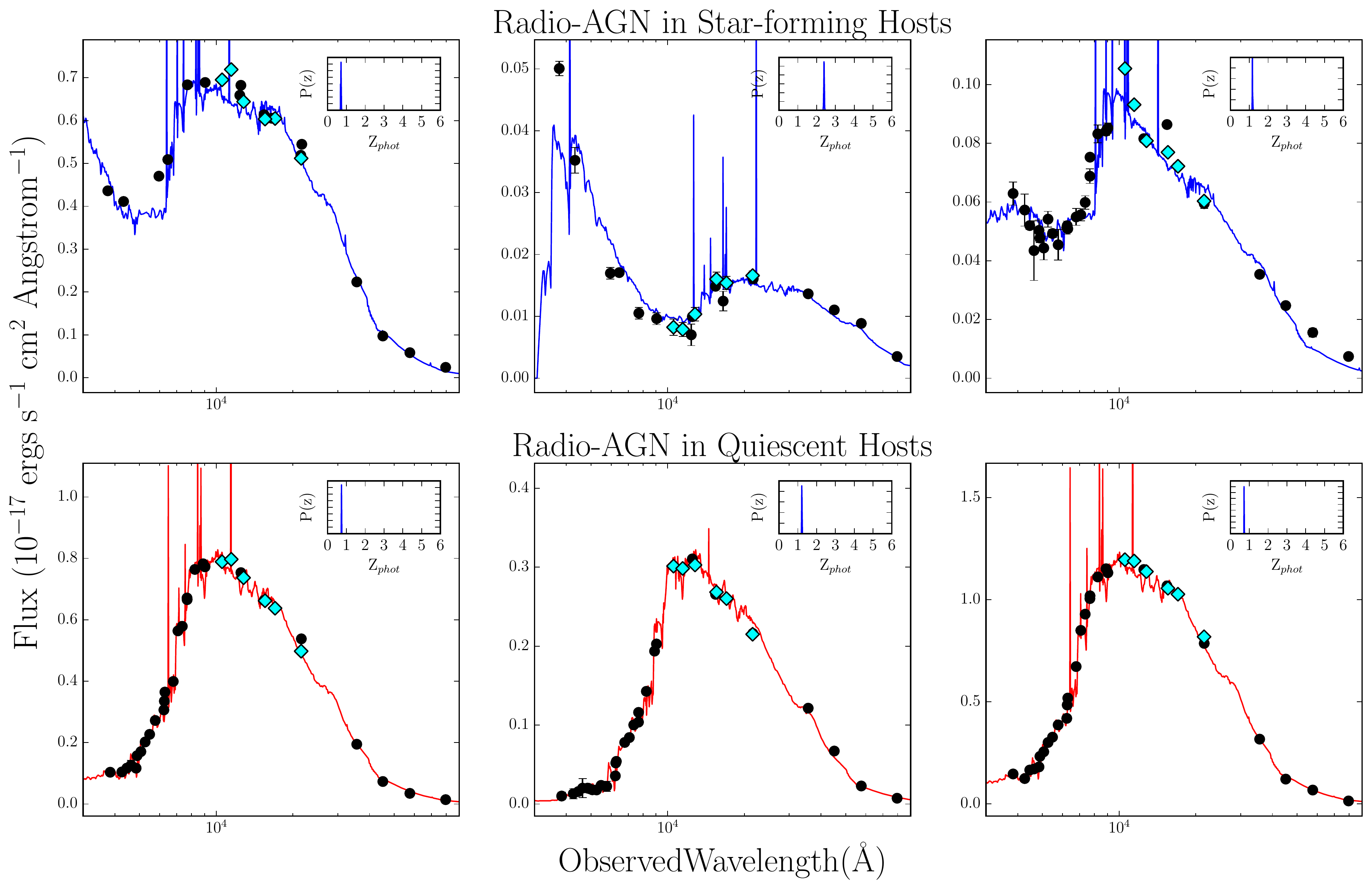}
	\caption{Example SED fits for two of our radio-detected populations: radio-AGN with quiescent hosts and radio-AGN with star-forming hosts (defined below). Photometric data points from ZFOURGE are shown (light blue) along with additional photometric data from other surveys used in the fitting process (black), error bars are show. The best fit EAZY SEDs are shown, colour-coded by their UVJ colour-code classification: quiescent (red) or star-forming (blue). On each panel a sub-plot displays the EAZY calculated probability distribution across redshift with narrow P(z) functions (in blue) indicating reliable photometric redshift fits. We can see that both types of radio-AGN hosts show good photometric fits and note that this is representative of the samples in general. SPITZER:MIPS 24$\mu$m fluxes were not used in the SED fitting process. }
	\label{fig:SEDs}
\end{figure*}	

Our primary data are the pre-release {\it Ks}-selected galaxy catalogues from the FourStar Galaxy Evolution Survey (ZFOURGE, Straatman et al, in prep, \citealt{Spitler2012, Tilvi2013, Straatman2014}). This survey covers the CDF-S (Chandra Deep Field South) and COSMOS (Cosmic Evolution Survey) fields \citep{Giacconi2001, Schinnerer2004}. Each 11~$\arcmin$ $\times$11~$\arcmin$ field is imaged at a resolution of 0.6~$\arcsec$ down to a 80\% {\it Ks}-band, point source magnitude limit of 24.53 and 24.74 AB mag for the CDFS and COSMOS fields, respectively. This corresponds to a measured 0.6"-radius aperture limit of 24.80 and 25.16 AB mag (5 sigma) in each field \citep{2014arXiv1412.3806P}. The inclusion of five near-infrared medium band filters in {\it J} and {\it H} ({\it J1}, {\it J2}, {\it J3}, {\it Hs} and {\it Hl}) along with deep multi-wavelength data from surveys such as CANDELS \citep{Grogin2011} allows the calculation of high-quality photometric redshifts using the photometric redshift code EAZY \citep{Brammer2008} which  uses a linear combination of 7 templates to produce the best fit SED. Additional 24$\mu$m observations from SPITZER:MIPS \citep{Rieke2004} are used to calculate star-formation rates but are not used in the EAZY SED fitting process. \\

Due to the relatively small volume probed by ZFOURGE, we also make use of the less sensitive, larger area Newfirm Medium Band Survey (NMBS, \citealt{Whitaker2011}) to increase the number of galaxies detected. NMBS covers an area of 27.6~$\arcmin$ $\times$ 27.6~$\arcmin$ in the COSMOS field and has significant overlap with the ZFOURGE COSMOS observations. While less sensitive in the {\it Ks}--selection band, NMBS still achieves a 5-sigma total magnitude of 23.5 in {\it K}--band and utilises up to 37 filters to produce accurate photometric redshifts ($\sim$1\%, Figure \ref{fig:ZphotZspec}: Left) for approximately 13,000 galaxies. \\

Overall, ZFOURGE and NMBS produce redshifts with errors of just 1\%-2\% (Normalised Median Absolute Deviation) \citep{Tomczak2014, Yuan2014, Whitaker2011} and are comparably accurate for radio-detected objects (Figure \ref{fig:ZphotZspec}: Left). However, radio-detected objects show a higher catastrophic failure rate from template mismatches and SED variability, which may be due to significant contamination of the stellar light by AGN emissions. Indeed, for the 6 radio-detected photometric redshift outliers, 2 objects have the smooth powerlaw SED shape associated with optical-AGN, 1 shows strong variability in the near-infrared (indicating potential IR-AGN activity) and the remaining 3 have poorly fitted SEDs due to coverage issues in some bands. These objects are removed during our flagging stages. \\

Figure \ref{fig:ZphotZspec}: Center shows a photometric versus spectroscopic redshift analysis limited to only those objects with detections in both surveys. Here we can see good agreement and comparable accuracy between ZFOURGE and NMBS out to z $\sim$ 1.0. The absence of high redshift spectra means we cannot directly test the accuracy of our photometric redshifts at values much higher than z $\sim$ 1.0, but we draw attention to the fact that both ZFOURGE and NMBS are specifically designed to produce accurate redshifts in the 1.5 $<$ z $<$ 2.25 regime for galaxies with strong 4000$\AA$/Balmer breaks. The majority of our z $>$ 1.0 radio galaxies have red SEDs with strong breaks similar to those shown in Figure \ref{fig:SEDs} and overall are well fitted by the stellar templates. Finally we note that the photometric redshifts from both surveys show good agreement with spectroscopic redshifts out to z $\sim$ 1.5 and with each other out to z $\sim$ 3 (Figure \ref{fig:ZphotZspec}: Right). \\

\begin{figure*}
	\centering
	\includegraphics[width=2.0\columnwidth]{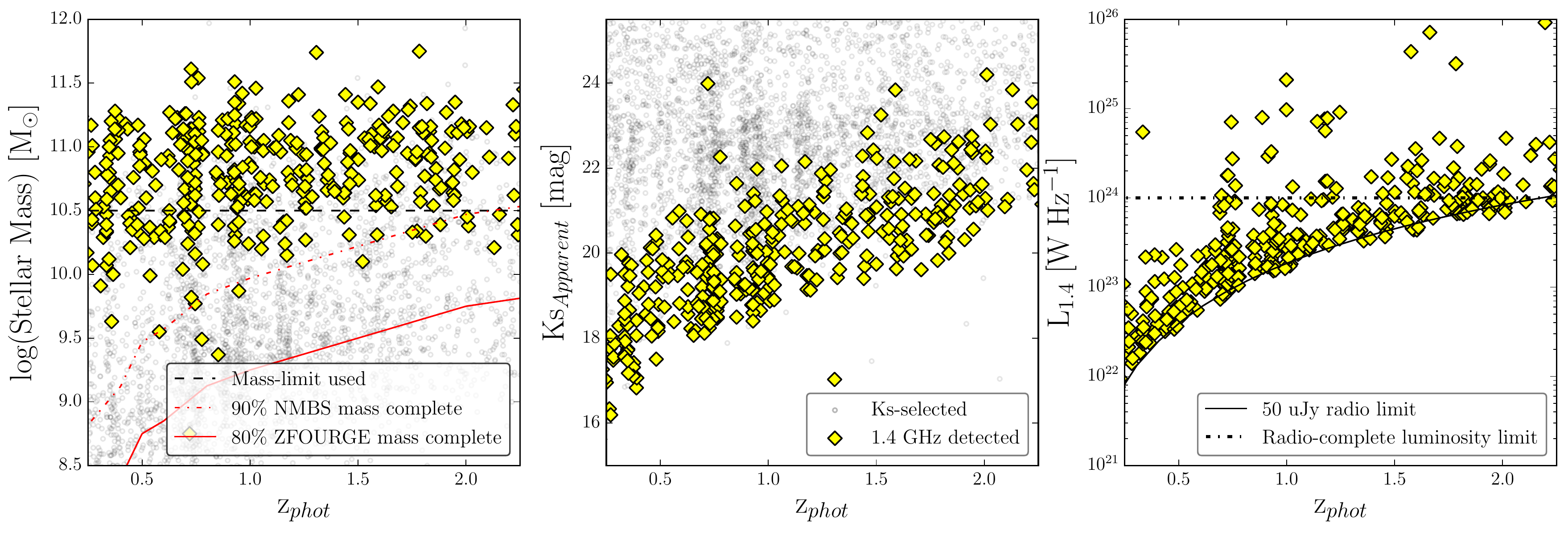}
	\caption{Stellar-mass, {\it Ks} magnitude and 1.4~GHz luminosities against z$_{phot}$ for the combined ZFOURGE and NMBS fields. The entire {\it Ks}-band selected sample is shown (black circles), alongside the 424 {\it Ks}-band objects with a 1.4~GHz counterpart within 1~$\arcsec$ (Yellow Diamonds). Left: The mass distribution of our galaxy sample shows a clear bias for radio detections to be found in massive galaxies. To account for this mass bias and to allow ready comparisons between radio detected and radio non-detected sources we define our mass-limited sample as objects above 10$^{10.5}$ M$_{\odot}$ (red dashes) and note that this is above both the 80\% ZFOURGE mass-completeness limit (red line, \citealt{2014arXiv1412.3806P}) and the 90\% NMBS mass completeness limit, which is the ZFOURGE 90\% curve scaled to match the quoted z=2.20 (red dot dashed line, \citealt{Wake2010}) out to our redshift limit of z = 2.25. Centre: The {\it Ks} magnitude distribution as a function of redshift for 1.4~GHz radio detections. Right: 1.4~GHz luminosities as a function of redshift along with our luminosity completeness limit of 10$^{24}$ W~Hz$^{-1}$ out to a redshift of z = 2.25. Note, for clarity only 10\% of the radio non-detected sources are plotted.}
	\label{fig:Mass}
\end{figure*}

	\subsection{1.4~GHz radio observations}

Radio detections in the CDFS field are determined from the images and catalogues of the Very Large Array (VLA) 1.4~GHz Survey of the Extended Chandra Deep Field South: Second Data Release \citep{Miller2013}. Covering an area approximately a third of a square degree with a average RMS of 7.4~$\mu$Jy per beam, this survey covers the entirety of ZFOURGE {\it Ks}-band observations in the CDFS and has an excellent angular resolution of 2.8$\arcsec$ by 1.6$\arcsec$. \\

Radio detections for the COSMOS field are taken from the 1.4~GHz VLA Cosmos Deep Project \citep{Schinnerer2010} which has an RMS of 10~$\mu$Jy per beam. The COSMOS Deep Project covers the entirety of the ZFOURGE and NMBS {\it Ks/K}-band observations with a angular resolution of 2.5~$\arcsec$ $\times$ 2.5~$\arcsec$. \\

\subsection{Cross-matching radio and K-band catalogues.} 
	\label{Sec:Cross} 

To check the astrometry between radio and {\it Ks} images we cross-match our catalogues using a 5$\arcsec$ separation radius and calculate the average positional offset. The COSMOS \& CDFS ZFOURGE images are found to have systematic astrometric discrepancies of 0.15$\arcsec$ and 0.30$\arcsec$ respectively when compared to the 1.4~GHz observations. The NMBS data shows only a 0.08$\arcsec$ offset from the COSMOS 1.4~GHz field. We apply corrections for these offsets in all subsequent cross-matching.\\

We now determine the separation at which we will assume radio and {\it Ks} sources are associated by calculating the number of radio-{\it Ks} pairs as a function of angular separation (0$\arcsec$--5$\arcsec$) for both NMBS and ZFOURGE. \\

To remove the effects of chance alignments, we create a model of randomly positioned, point like sources and subtract this model from the observed distribution. The resultant distribution is largely Gaussian with 68\% of radio-{\it Ks} pairs separated by less than 0.32$\arcsec$ (Figure \ref{fig:Seps}) for ZFOURGE and 0.20$\arcsec$ for NMBS. Due to the quoted 0.3$\arcsec$ accuracy of the \citet{Miller2013} radio catalogues we therefore adopt a cross-matching radius of 1$\arcsec$ for all fields. \\

Using this limit we cross-match our radio catalogues against {\it Ks/K} band sources with ZFOURGE and NMBS use-flag of 1 which ensures that they have sufficient optical and IR imaging coverage, are far away from the scattered light of bright stars and are not stars themselves. We find 546 radio sources (ZFOURGE: 160 and NMBS: 386) have a {\it Ks}-detected counterpart and note that our model for random alignments predicts that less than 14 of these are due to chance (ZFOURGE: 4 and NMBS: 10). \\

A visual inspection of the cross-matching shows that 2 objects in the COSMOS field seem to be missed in our automated procedure, despite being well aligned with the head of a head--tail radio galaxy and the core of a Faranoff-Riley Type I radio jet \citep{Fanaroff1974}. These objects were missed because our selected radio-catalogues quote the geometric center of large, extended objects. We manually add these matches to both the ZFOURGE and NMBS COSMOS sample bringing our total up to 550 (ZFOURGE: 162 and NMBS: 388) radio-{\it Ks} pairs. Thus we detect 550 radio-objects in the ZFOURGE and NMBS area, or approximately 91\% (550/603) of radio objects that have good coverage in all bands and are not flagged as near stars. The remaining 9\% of sources show no {\it Ks}-band counterpart within 1 arc-second and are potentially {\it K}-band drop-outs. We include a list of these objects in Appendix Table 3. \\

\begin{table}
\label{tab:Flags}
\centering
\caption{Flagging breakdown, the number of sources remaining in the catalogues after applying each flagging stage.}
\raggedright
\begin{tabular}{@{}p{3.5cm}p{2cm}p{2cm}@{}}
Radio Sources & NMBS & ZFOURGE\\
\hline
With \it{Ks}-counterpart & 388 (91+-2\%) & 162 (89+-2\%) \\
Post F$_{1.4}$  $<$ 50$\mu$Jy flag & 381	& 115 \\
Post duplicate flag	& 317 & 115\\
Post SED flag & 312 & 107\\
\hline
\end{tabular}
\end{table}

We further flag our sample to ensure sensible handling of the overlapping NMBS and ZFOURGE fields, to minimize the difference in radio sensitivity between fields and to catch any poor SED fits missed by the automated procedure (Table 1). \\ 

To do this, we flag all objects with 1.4~GHz fluxes less than 50$\mu$Jy in order to keep our radio sensitivity constant across all fields. This reduces our radio sample down to 496 objects (ZFOURGE: 115 and NMBS: 381).\\

Some NMBS and ZFOURGE sources are the same due to the overlapping observations. Inside the overlapping region between NMBS and ZFOURGE we preferentially use ZFOURGE data wherever possible as the greater signal to noise should result in more secure redshift fitting for dusty star-forming and quiescent objects (which dominate our sample) at redshifts greater than 1.5. If however, any radio-detection does not have a ZFOURGE counterpart, for example due to flagging or sensitivity issues, it will have been included in the analysis wherever possible via the NMBS observations. Outside the overlapping region we use NMBS exclusively. After duplicate flagging there remains 432 objects with reliable {\it Ks/K} and radio band detections shared between ZFOURGE and NMBS (Appendix Tables 1 and 2). The {\it Ks/K}-band, radio luminosity and stellar-mass distributions of this sample are shown in Figure \ref{fig:Mass}. \\

Finally we manually flag 13 objects due to poorly fit SEDs. Two of these objects objects are flagged due to saturated or missing coverage that is missed by the automated flagging due to the averaging of sensitivity coverage over 1.2 by 1.2 arc-second areas. Of the remaining 11 sources, 7 possess pure power-law SEDs whose redshift estimate cannot be trusted as there is no break in the SED for EAZY to fit too and 4 SEDs showing large variability between observations in the NIR bands. Hence we have flagged $~$2.5\% of our final radio sample due to poor SED fits. \\
 
\section{Classifying radio galaxies}
\label{Sec:Class}

\subsection{By Host Type}
    \label{Sec:UVJ}
  
\begin{figure}
	\centering
	\includegraphics[width=1.0\columnwidth]{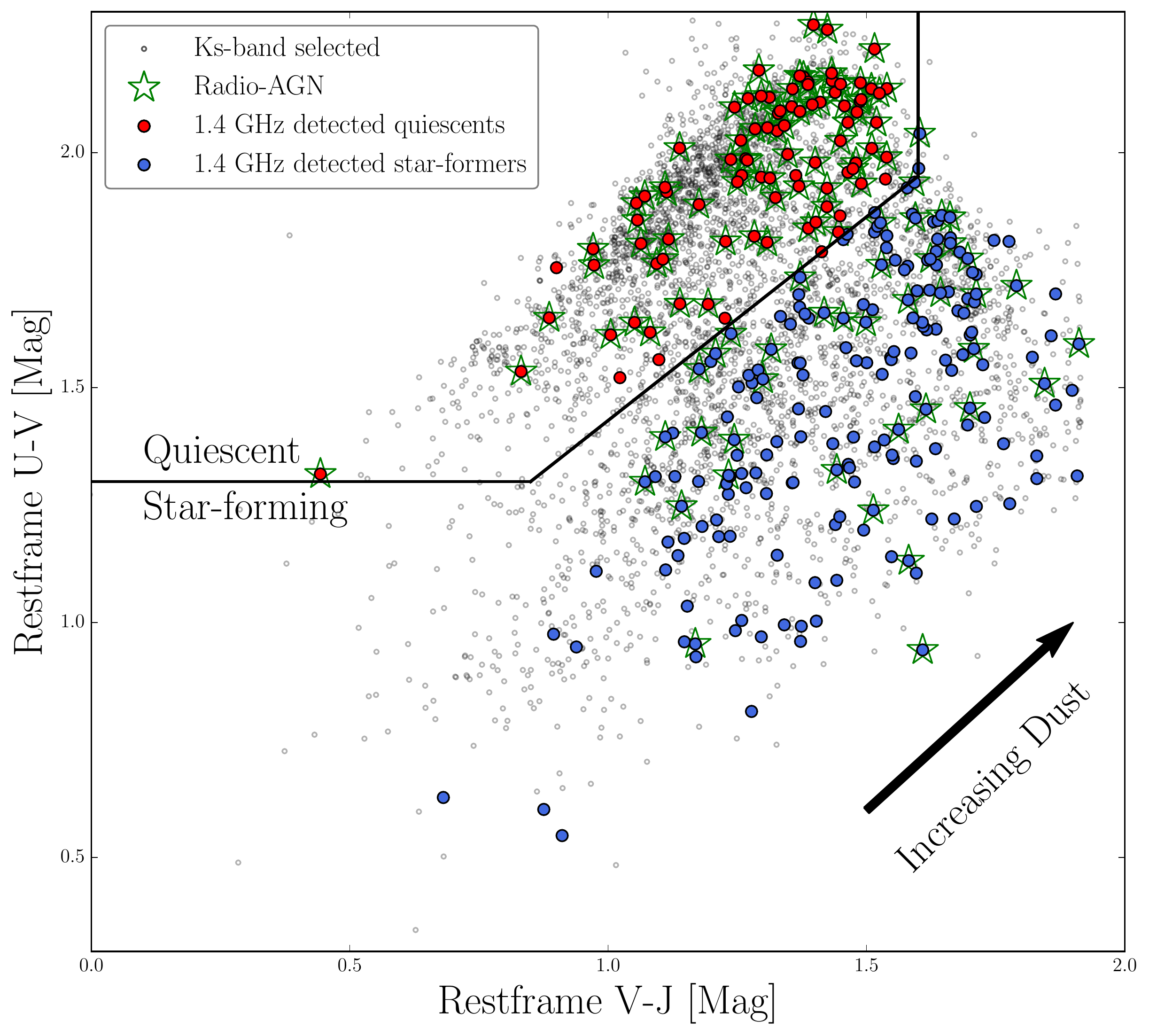}	
    \caption{The rest-frame, UVJ colour-colour diagram for radio-detected star-formers (blue circles), radio-detected quiescent galaxies (red circles) and the ZFOURGE {\it Ks}-band mass-limited sample (black circles). Radio-AGN are selected based on their Radio to UV+24$\mu$m SFR ratios and are highlighted (green stars). For clarity only 10\% of the radio non-detected sources are plotted.}
	\label{fig:UVJ-AGN}
\end{figure}

We determine the stellar population properties, including Mass, dust-content, rest-frame colors and luminosities of our sample by fitting \citet{Bruzual2003} stellar population synthesis models with FAST \citep{Kriek2009} using a \citet{Chabrier2003} initial mass function. Figure \ref{fig:SEDs} shows example SEDs and fits. Finally galaxies were divided into two stellar population types: quiescent or star-forming, based upon their UVJ position \citep{Wuyts2007}, shown in Figure \ref{fig:UVJ-AGN}. \\

\subsection{By Radio-AGN activity}
	\label{Sec:AGN}
\begin{figure*}
	\centering
	\includegraphics[width=2.0\columnwidth]{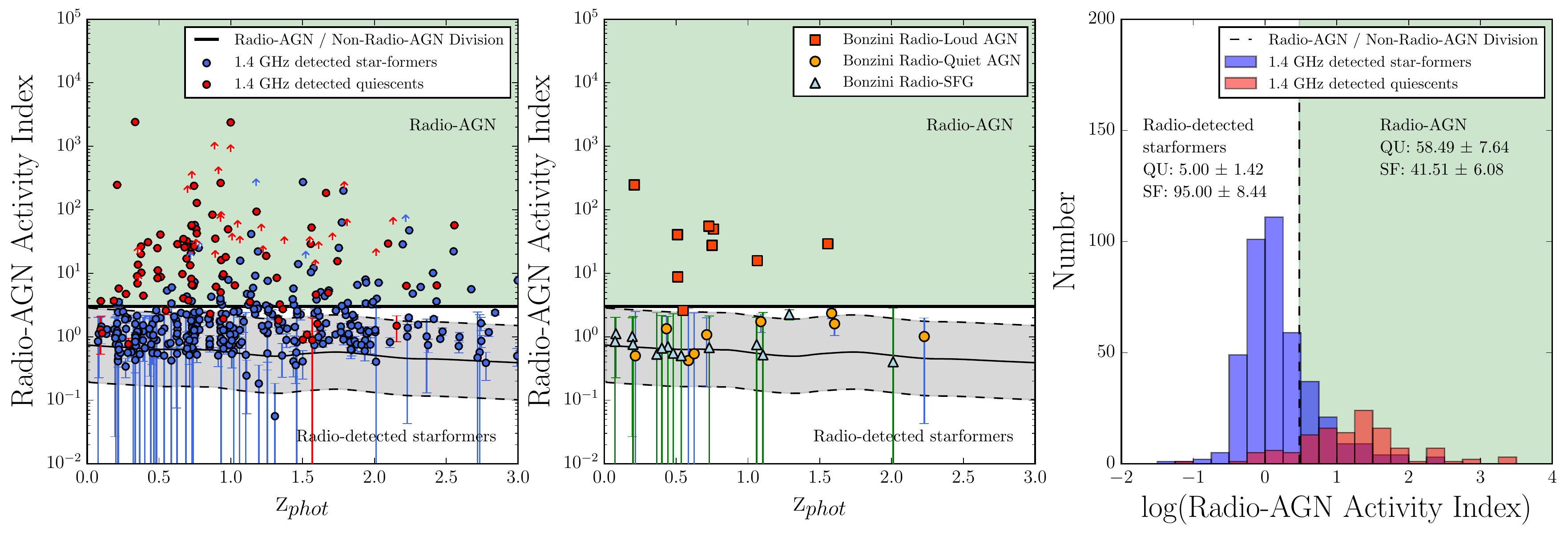}\\
	\caption{Left: the ratio of UV+IR to Radio SFR as a function of redshift, colour-coded by stellar population type. Objects with 24~$\mu$m fluxes less than the 24~$\mu$m sensitivity are shown as 3--sigma lower limits (arrows). The \citet{Wuyts2008} average star-forming galaxy SED template is shown (thin black line) along with the three sigma scatter seen the 1.4~GHz to 24$\mu$m flux ratios of local star-forming galaxies (grey shaded area). We determine that objects above the upper limit of this line (i.e  SFR$_{1.4 GHz}$ / SFR$_{UV+IR}$ $>=$ 3.0) are considered to be radio-AGN (green shading). Centre: A comparison of our classification of radio-AGN and non-radio-AGN sources to the \citet{Bonzini2013} classification scheme: Radio-Loud AGN (dark orange squares), Radio-Quiet AGN (light orange circles) and radio detected star-forming galaxies (light blue triangles). We see good agreement between Bonzini Radio-Loud AGN and our radio-AGN sample. Right: A breakdown of combined UV+24$\mu$m to radio star-formation rate ratios as a function of quiescent (red) or star-forming (blue) host galaxies. The breakdown of host types for radio-AGN (green shaded) and radio detected star-formers (white shaded) is labelled.}
	\label{fig:Bonzini-Test}
\end{figure*}

A large number of techniques are currently used to identify radio-AGN (see \citealt{Hao2014b} for a review of 9 separate methods) but a key indicator is the ratio of 24$\mu$m to 1.4~GHz flux or the ratio of FIR (70$\mu$m-160$\mu$m) to 1.4~GHz flux. The resultant tight linear correlation shows no evolution out to a redshift of two \citep{Mao2011} and is commonly known as the far-infrared radio correlation. Objects lying significantly off this correlation are typically classified as radio-AGN but the offset required for this depends on the AGN SED model and redshift. \\ 

It has been suggested that low redshift, radio-quiet AGN are dominated in radio wavelengths by host galaxy star-formation \citep{Bonzini2013, Kimball2011} and composite sources in which AGN and star-formation both contribute to the radio emission are well--known \citep{Emonts2008, Emonts2014, Carilli2013}. We therefore define a radio-AGN as an object in which the total radio flux is significantly greater than the radio emission expected from the star-formation rate determined at other wavelengths.  \\

The baseline star-formation rate of each object in our sample is calculated using a combined optical and infrared technique \citep{Bell2005, KennicuttJr.1998} and utilises the same initial mass function \citep{Chabrier2003} as used in the galaxy mass fitting process.

\begin{equation}
	\centering
	SFR_{UV+FIR} = 1.09 \times (L_{IR} + 3.3 \times L_{2800}) ~ [M_{\odot} {\rm yr^{-1}}]
	\label{Eq:IRBell}
\end{equation}

where L$_{IR}$ is the bolometric rest-frame infrared luminosity between 8 and 1000$\mu$m. This is extrapolated from SPITZER:MIPS 24$\mu$m fluxes \citep{Rieke2004} using the \citet{Wuyts2011} average stellar template. L$_{2800}$ is the EAZY interpolated rest-frame 2800 $\AA$ (UV) luminosity. \\

This SFR allows for the accurate tracing of star-formation in a non-dusty host via UV emissions from hot young stars and complements this with the contribution from dust obscured stars using 24~$\mu$m observations. \\

Our radio star-formation rate estimate is just the linear, high luminosity portion of the \citet{Bell2003} SFR estimate scaled to a Chabrier initial mass function \citep{Karim2011} as all of our sources are above the given critical radio luminosity (6.4 $\times$ 10$^{21}$, W~Hz$^{-1}$):

\begin{equation} \label{Eq:RadioBell}
SFR_{Radio} =   3.18 \times 10^{-22} L_{\rm 1.4\,GHz} ~ [M_{\odot} yr^{-1}]
\end{equation}

Where L$_{1.4}$ is calculated using a radio spectral index $\alpha =$ -0.3 (where S $\propto \nu^{\alpha}$ in order to match the spectral index of the radio end of the Wuyts template used for calculating UV+IR SFR. Radio based star-formation rates are also largely unaffected by dust obscuration. Hence, they should accurately match the combined UV+IR based estimates within errors unless extra emissions from radio-AGN are present. Figure \ref{fig:Bonzini-Test}: Left shows our method for identifying radio-AGN using their ``Radio-AGN Activity Index":

\begin{equation} \label{Eq:Activity Index}
$$Radio-AGN Activity Index =$$ SFR_{Radio} / SFR_{UV+IR}
\end{equation}

We use the \citet{Wuyts2011} average star-forming template to predict the expected indexes of star-forming galaxies. The Wuyts template produces a nearly flat, ratio of SFRs which is only slightly dependant upon redshift, hence we approximate the dividing line between star-formation driven radio emission and AGN driven radio emission as the three times the scatter seen in local 1.4~GHz to far-infrared flux ratios (0.39 dex, \citealt{Moric2010}). This divides our radio-detected sample into two: objects whose radio-AGN Activity Index is less than 3.0 are henceforth classified as radio-detected star-formers and those whose radio-AGN activity index is greater than 3.0 are classified as radio-AGN. Figure \ref{fig:Bonzini-Test} (Center) compares our radio-AGN selection criteria to that of \citet{Bonzini2013}. We see good agreement between our radio-AGN sample and their sample of radio-loud AGN. Finally we note that our results do not change significantly if our radio-AGN selection criteria is raised from a Radio-AGN Activity Index of 3.0, up to a Index of 5.0. The breakdown of quiescent and star-forming hosts as a function of our radio-AGN selection criteria is shown in Figure \ref{fig:Bonzini-Test} (Right). \\

While the combined UV+IR SFR should give a good estimate of total star-formation rate, independent of dust obscuration for the majority of sources, it does however, assume that the entirety of the observed 24$\mu$m flux is from star-formation and not contaminated by a hot, dusty AGN torus. To determine how much of an impact this issue will have on our analysis, we apply the Donley color wedge \citep{Donley2012} to our final analysis sample to determine the the number of radio sources with infrared-AGN colours. We find that 10$\pm$2\% (40/412) radio-detected objects in our sensitivity limited sample have IR-AGN colors and note that these objects may have artificially elevated SFRs. \\

But how does AGN contamination of the 24$\mu$m flux affect our analysis of SSFRs in radio-AGN and mass-similar galaxies? Of the 40 IR-AGN identified, only 10 fall within our final mass and radio luminosity complete sample of 65 objects (15$^{+5}_{-3}$\%) and only 4 of these are also classified as radio-AGN (of which there are 42). If we remove these objects from our radio-AGN analysis, along with every identified IR-AGN candidate from the mass and luminosity complete control sample of non-radio-AGN, we find no significant change in our findings. \\

Due to the presence of IR-AGN it is also possible that we mis-classify a fraction of radio-AGN as radio-star-formers due to contamination of the 24$\mu$m flux. Any infrared flux from an AGN will artificially inflate the host galaxies SFR$_{UV+IR}$ estimate, potentially keeping it below the division in Figure \ref{fig:Bonzini-Test} (Left). Because of this, it is possible that with better FIR observations some sources may move upwards from their current positions in Figure \ref{fig:Bonzini-Test} (Left) and become re-classified as radio-AGN. To quantify the how common these objects are, we calculate the number of IR-AGN in the mass and radio luminosity complete sample that are not currently classified as radio-AGN. It is possible that these 6 objects should be included in our radio-AGN sample. Therefore, we estimate that our Radio-AGN sample is at worst 88$^{+3}_{-6}$\% (42/(42+6)) complete due to this effect and note that the inclusion of these objects to our analysis makes no significant changes to our findings.\\

To account for these effects, the upper error bar on the quoted Radio-AGN fraction for the sensitivity limited sample given in Section \ref{Sec:RadSky} includes the value we attain should all 40 of the IR-AGN in the sensitivity limited sample currently identified as radio-detected star-formers were to move into our radio-AGN sample. The equivalent has been done for our AGN fraction versus redshift analysis shown in Figure \ref{fig:AGNFrac}.


\section{Analysis}
	\label{Sec:Analysis}
	\subsection{The Radio sensitivity limited sky}
	\label{Sec:RadSky}

	The upcoming deep radio survey The Evolutionary Map of the Universe (EMU), will observe approximately 75\% of the radio sky down to a limiting 5-sigma sensitivity of 50~$\mu$Jy and produce a catalogue of approximately 70 million radio galaxies \citep{Norris2011}. To prepare for this large undertaking, we now characterise the radio-sky at a comparable sensitivity to EMU.\\

	The majority of radio objects (76\%) are found to have rest-frame UVJ colors associated with actively star-forming galaxies. Notably, the star-forming radio population shows a significant bias towards high dust content, with V-J values typically above the dusty, non-dusty border at V-J = 1.2 \citep{Spitler2014}. \\
 
	Radio sources are found to have a median mass of 10$^{10.8 \pm 0.5}$~M$_{\odot}$, a median redshift of z = 1.0$\pm$0.7, a median SSFR of 0.8$\pm$3.7~M$_{\odot}$~Gyr$^{-1}$ and a median visual extinction of 1.4$\pm$0.9~mag. The errors quoted here are the standard deviations as we are characterising the populations rather than constraining the medians. We determine that only 38 $\substack{+6 \\ -2}$\% of radio sources (1-sigma BETA confidence interval error, \citep{Cameron2013}) above 50~$\mu$Jy are radio-AGN and this fraction shows no evolution within errors in the redshift range of 0.25 to 2.25. This confirms the result by \citet{Seymour2008} that the sub-mJy radio sky is dominated by emissions from star-formation.  \\
\begin{table*}
\label{Tab:SD-Rs}
\centering
\begin{tabular}{@{}cccccc@{}}
\multicolumn{4}{c}{{\bf Table 2.}}\\
\hline
Property & Radio-detected & Mass-Similar & KS P-Value  \\
\hline 
log(Mass) [M$_{\circ}$] & 11.03 $\pm$ 0.06 & 11.00 $\pm$ 0.02 & P = 0.98 \\
Redshift & 1.68 $\pm$ 0.06 & 1.54 $\pm$ 0.16 & P = 0.10 \\
SSFR [M$_{\circ}$ Gyr$^{-1}$] & 3.92 $\pm$ 0.59 & 0.78 $\pm$ 0.21 & P = 0.00 \\
Visual Extinction [Mags] & 1.60 $\pm$ 0.14 & 1.40 $\pm$ 0.19 & P = 0.36 \\
\hline
\end{tabular}
\caption{Median properties for the 23 ``radio-detected star-former" galaxies (Median L$_{1.4 GHz}$ = 1.34 $\times$ 10$^{24}$ W~Hz$^{-1}$) in ZFOURGE and NMBS, compared against the mass-similar, redshift binned radio non-detected star-former sample. }

\label{Tab:AGNQU}
\centering
\begin{tabular}{@{}cccc@{}}
\multicolumn{4}{c}{{\bf Table 3.}}\\
\hline
Property & Radio-AGN (QU) & No Radio-AGN (QU) & KS P-Value \\
\hline 
log(Mass) [M$_{\circ}$] & 11.14 $\pm$ 0.06 & 11.12 $\pm$ 0.02 & P = 0.98 \\
Redshift & 1.10 $\pm$ 0.13 & 0.99 $\pm$ 0.16 & P = 0.56 \\
SSFR [M$_{\circ}$ Gyr$^{-1}$] & 0.08 $\pm$ 0.13 & 0.06 $\pm$ 0.02 & P = 0.56 \\
Visual Extinction [Mags] & 0.55 $\pm$ 0.10 & 0.50 $\pm$ 0.09 & P = 0.82 \\
\hline
\end{tabular}
\caption{Median properties for the 22 radio-AGN in quiescent hosts (Median L$_{1.4 GHz}$ = 3.1 $\times$ 10$^{24}$ W~Hz$^{-1}$) in ZFOURGE and NMBS, compared against the mass-similar, redshift binned, non-AGN quiescent population.}

\label{Tab:AGNSF}
\centering
\begin{tabular}{@{}cccc@{}}
\multicolumn{4}{c}{{\bf Table 4.}}\\
\hline
Property & Radio-AGN (SF) & No radio-AGN (SF) & KS P-Value \\
\hline 
log(Mass) [M$_{\circ}$] & 10.95 $\pm$ 0.09 & 10.92 $\pm$ 0.03 & P = 0.97 \\
Redshift & 1.64 $\pm$ 0.12 & 1.55 $\pm$ 0.16 & P = 0.50 \\
SSFR [M$_{\circ}$ Gyr$^{-1}$] & 1.41 $\pm$ 0.96 & 0.79 $\pm$ 0.25 & P = 0.13 \\
Visual Extinction [Mags] & 1.45 $\pm$ 0.16 & 1.45 $\pm$ 0.21 & P = 0.77 \\
\hline
\end{tabular}
\caption{Median properties for the 20 radio-AGN in star-forming hosts (Median L$_{1.4 GHz}$ = 2.7 $\times$ 10$^{24}$ W~Hz$^{-1}$) in ZFOURGE and NMBS, compared against the mass-similar, redshift binned, non-AGN star-forming population.}

\end{table*}

	\subsection{The Mass-limited and Mass-Similar, Radio-Complete sky}
	\label{Sec:Sample}
	We now limit our sample in terms of mass, radio-luminosity and redshift to allow for a fair comparison between high-mass radio galaxies and their non-radio counterparts using the following criteria:
	\begin{enumerate}
	\item 0.25 $\le$ z $<$ 2.25 to minimize the effects of our small survey volume and ensure completeness in both radio-luminosity and mass. \\	
	\item Radio luminosities greater than 10$^{24}$ WHz$^{-1}$ at 1.4~GHz observed. Objects above this limit are referred to hereafter as high-luminosity sources and objects below this limit as low-luminosity sources.
	\end{enumerate}
	
	In addition we produce two slightly different control samples: 
	\begin{enumerate}
	
	\item The ``mass-similar" population is used for comparing the properties of our test groups (both radio-AGN and radio-detected star-forming galaxies) to a control sample comprised of galaxies with similar redshifts and mass. In each case we determine the number of objects within our test population and randomly select the same number of {\it Ks}-detected galaxies from within each of our redshift bins (0.25-1.00, 1.00-1.65 and 1.65-2.25) whose stellar-masses are within 0.1 log(M$_{\odot}$) of an object in the test group. We then measure and record the median value for each property of interest (e.g. SFR) for this mass-similar control population and repeat the process 1000 times. The median value of all these measurements is then compared to the median of the test population (radio-AGN or radio-detected star-formers). Errors on all test population properties are standard errors and errors on the mass-similar control samples are calculated using the normalised median absolute deviation of the 1000 random samples. \\
	\item The ``mass limited" population is simply all objects with stellar masses greater than 10$^{10.5}$ M$_{\odot}$ (Figure \ref{fig:Mass}: Left) and used solely for determining the fraction of galaxies containing a radio-AGN. \\ 
	\end{enumerate}
	
	Both our mass-similar and mass-limited samples are above the mass completeness limit for the both the ZFOURGE (80\% limit = 7.8 $\times$ 10$^{9}$ M$_{\odot}$, \citealt{2014arXiv1412.3806P}) and NMBS (90\% limit = 3.0 $\times$ 10$^{10.0}$ M$_{\odot}$, \citealt{Wake2010,Brammer2011}) surveys at our maximum redshift of z = 2.25. Finally, all percentage and fraction errors in this paper are 1-sigma values calculated using the BETA confidence interval \cite{Cameron2013}.

	\subsubsection{Radio-detected star-formers galaxies}

	We now consider radio galaxies whose Radio-AGN Activity Index is less than 3.0 (i.e. objects whose radio emissions are consistent with what we would expect to detect based on their star-formation rate). In Figure \ref{fig:UVJ-AGN} we see that these high-luminosity ``radio-detected star-formers" are associated with star-forming rest-frame UVJ color (91$^{+3}_{-10}$\%) and that the 9\% of these objects found in the quiescent UVJ region are located very near the quiescent--star-forming boundary. Of these 2 objects, we find that 1 has an X-ray detection in public catalogues. In general these objects are thought to have simply scattered outside the UVJ star-forming region, but it is also possible that we are seeing the last effects of residual star-formation in a largely quiescent host. \\

As a whole, we find high-luminosity radio-detected star-forming galaxies to possess higher  star-formation rates (P $<$ 0.01) than their mass and redshift similar star-forming counterparts (Table 2) and we note that the low KS-test P-value for this property is likely due to the sensitivity bias of our radio observations towards high-luminosity (and hence high star-formation rate) sources.

	\subsubsection{Radio-AGN}
	\label{Sec:RAGN}

\begin{figure}
	\centering
	\includegraphics[width=0.8\columnwidth]{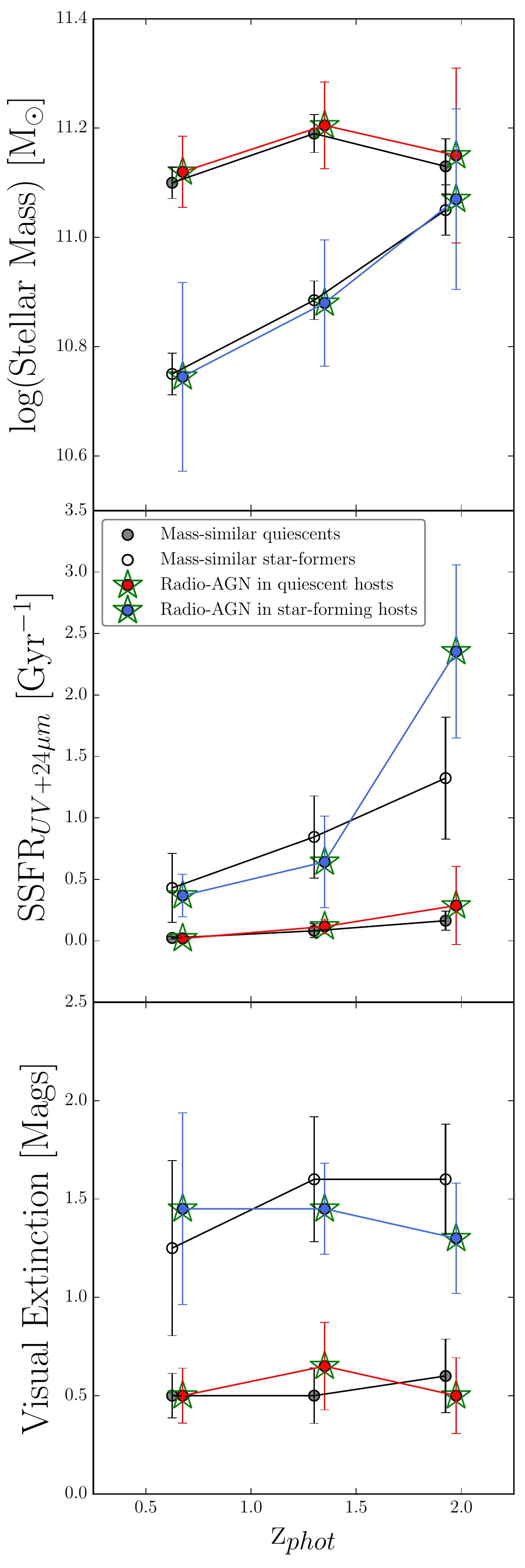} 
	\caption{The measured median values for Mass (top), SSFR (middle) and Visual Extinction (bottom) for high-luminosity, quiescent and star-forming radio-AGN compared to their mass-similar counterparts. We see no evidence for enhanced or suppressed star-formation in high-luminosity radio-AGN hosts within our redshift range. Errors are standard errors on the median and the normalised median absolute deviation, for radio-AGN and the mass-similar sample respectively. Values are offset by 0.10 in the horizontal direction for clarity.}
	\label{fig:AGNProps}
\end{figure}

\begin{figure}
	\centering
	\includegraphics[width=1.0\columnwidth]{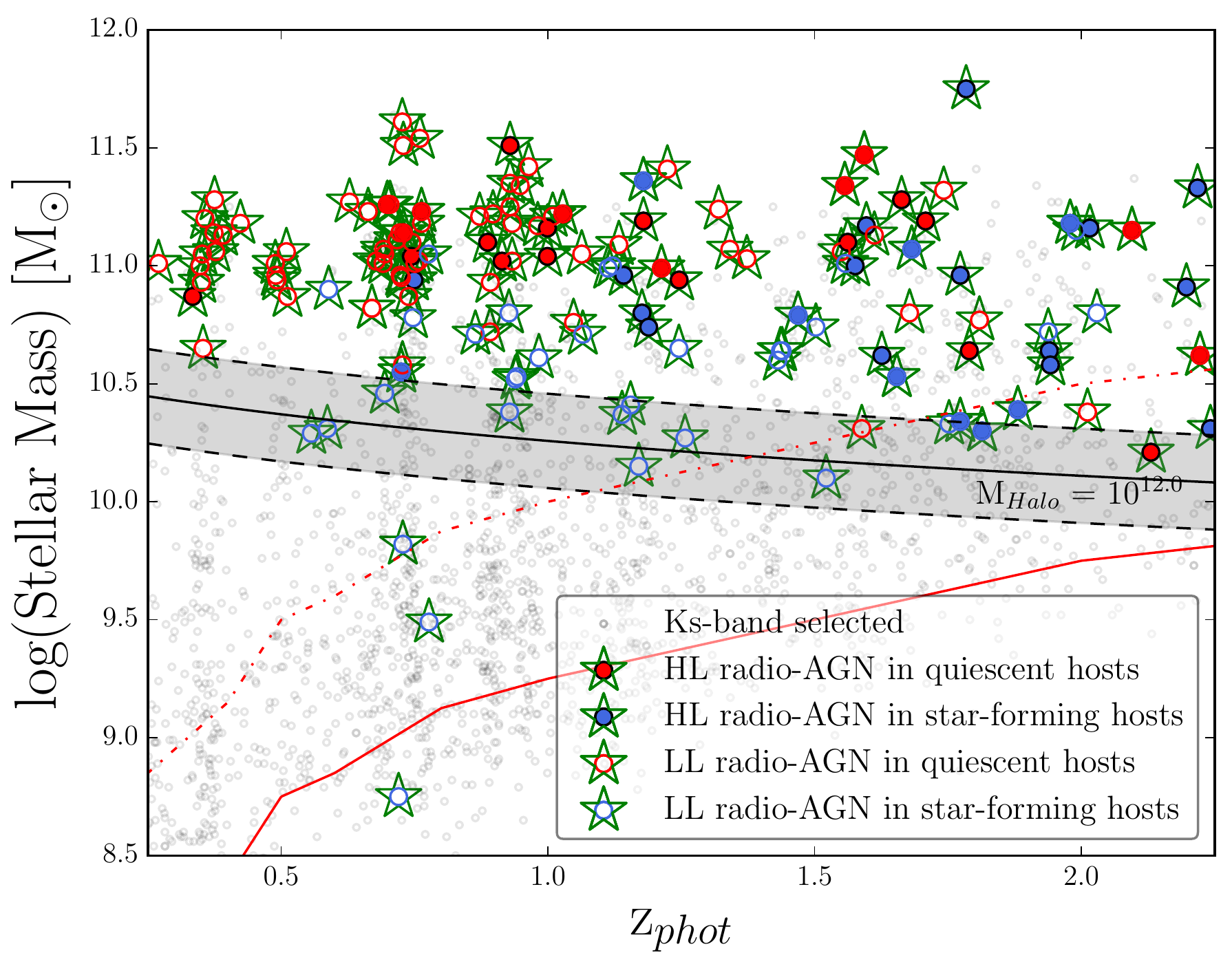} 
	\caption{The stellar-masses of high (L$_{1.4}$ $\ge$ 10$^{24}$ W~Hz) and low (L$_{1.4}$ $\le$ 10$^{24}$ W~Hz) luminosity radio-AGN. (filled and empty circles with green stars respectively). Quiescent and star-forming host types are also shown (red and blue circles respectively). The stellar-mass equivalent of a 10$^{12}$ M$_{\odot}$ halo-mass (black line) along with the {\it Ks}-selected sample (black circles) are plotted for reference, for clarity only 10\% of the {\it Ks}-selected sample is plotted. Finally, the 80\% ZFOURGE mass-completeness limit is shown (red line, \citealt{2014arXiv1412.3806P}) along with the 90\% NMBS mass completeness limit, extrapolated down from z=2.20 (red dot dashed line, \citealt{Wake2010}). Radio-AGN in quiescent hosts are only found in objects with halo masses above 10$^{12}$ M$_{\odot}$ (black line), 100\% of high-luminosity radio-AGN in our sample are found above this line. The 1-sigma local scatter in the stellar-to-halo mass relation is shown (grey shaded region). Finally we note that for this plot we do not limit our radio-AGN sample by mass in any way.}
	\label{fig:MassLim}
\end{figure}	

	\begin{figure}
\includegraphics[width=0.82\columnwidth]{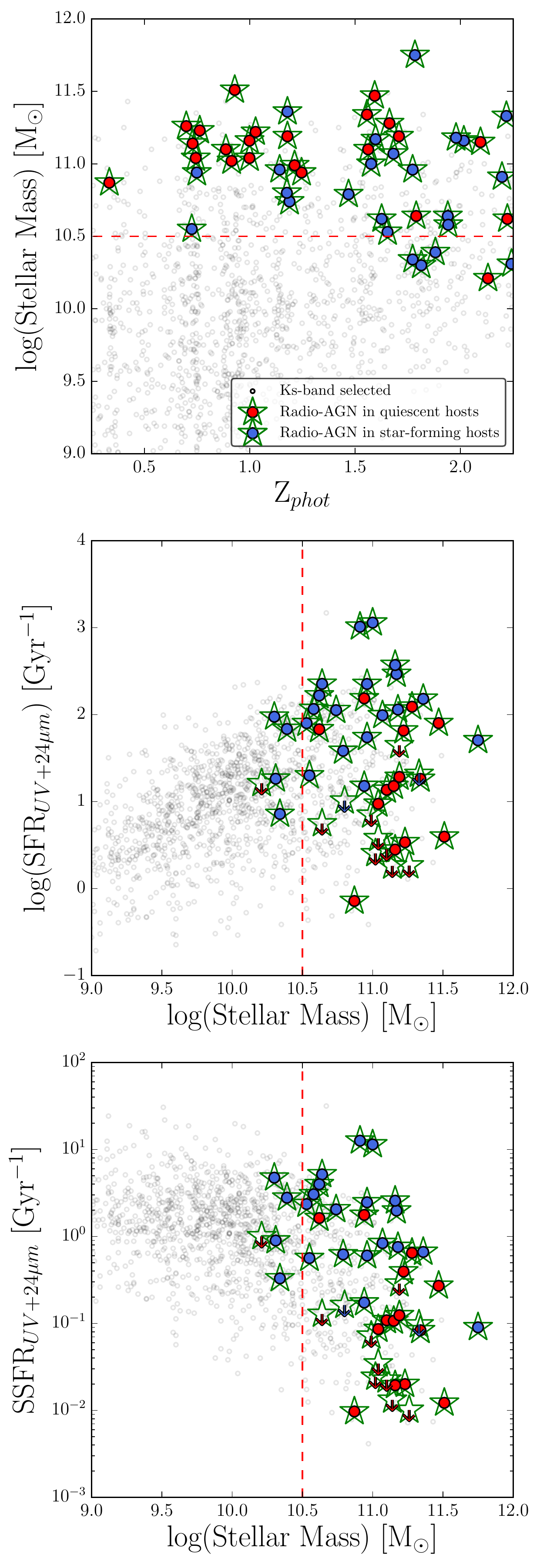}
	\caption{Top: The mass-redshift relation for high-luminosity radio-AGN in star-forming hosts (blue circles with green stars) and high luminosity radio-AGN in quiescent hosts (red circles with green stars). For reference, the {\it Ks}-selected, sample is also shown (grey circles) along with the mass-limit used for our analysis of the various populations  (red dashed line). For clarity only 10\% of the {\it Ks}-selected sample is plotted. Middle: The mass versus SFR plot colour-coded as before. Bottom: The mass vs SSFR plot colour-coded as before .}
	\label{fig:HybMass}
\end{figure}

While low-redshift radio-AGN are traditionally associated with massive, ellipticals \citep{Lilly1987, Owen1989, Vron-Cetty2001}, our sample shows a variety of rest-frame colors, with only 48\% $\pm$ 7\% of high-luminosity radio-AGN found within quiescent galaxies. The remaining 52\% $\pm$ 7\% are hosted in star-formers with high dust-contents which is in good agreement with the growing number of CO detections found in high-z radio-galaxies \citep{Emonts2008, Emonts2014, Carilli2013}. \\

Quiescent galaxy hosts with high-luminosity radio-AGN are found to possess star-formation rates and dust-contents indistinguishable from the mass-similar quiescent population (P = 0.56 and 0.82 respectively, see Table 2). With Figure \ref{fig:AGNProps} showing no evidence to suggest either suppressed or enhanced star-formation in our redshift range of 0.25 and 2.25. \\

To determine the evolving stellar (and corresponding fixed halo) mass at which high-luminosity radio-AGN appear, we plot the masses of radio-AGN as a function of redshift (see Figure \ref{fig:MassLim}). Comparing these observations to a fixed halo-mass, which has been converted to stellar-mass using the \citet{Moster2010} stellar-to-halo mass relation (Moster EQ: 24), we determine the halo-mass that bounds the high-luminosity radio-AGN population to be 10$^{12}$ M$_{\odot}$. \\

\begin{figure}
	\centering
	\includegraphics[width=1.0\columnwidth]{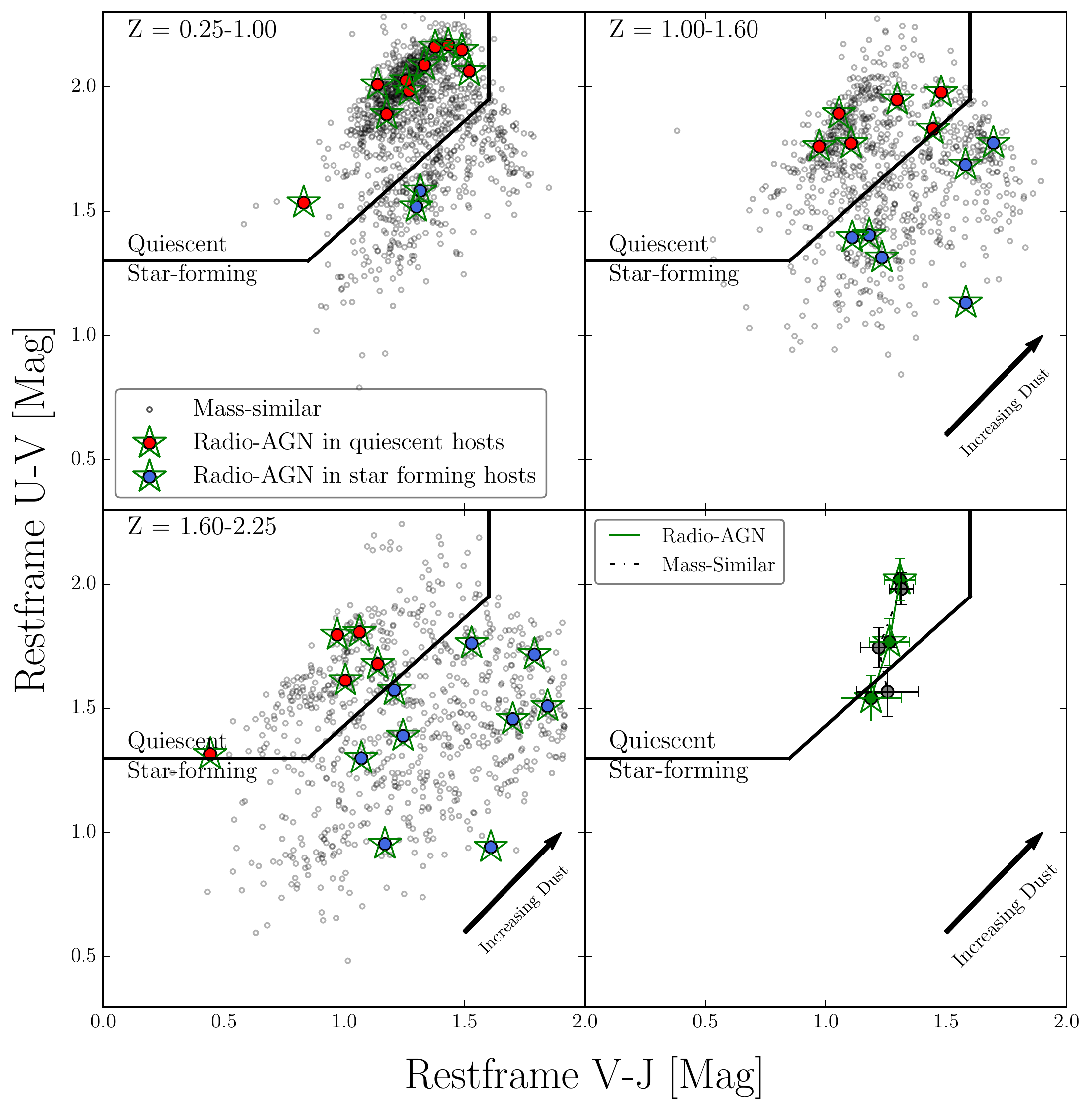}
	\caption{The UVJ diagram for high-luminosity radio-AGN compared to the mass-similar sample in three redshift bins: 0.25-1.0 (top left), 1.0-1.60 (top right) and 1.60-2.25 (bottom left). Bottom right: The median position for radio-AGN and the randomly sampled mass-similar population across the three redshift bins. Errors are standard error on the median and the Normalised median absolute deviation respectively. For clarity only 10\% of the {\it Ks}-selected sample is plotted}
	\label{fig:AGNUVJ}
\end{figure}

\begin{figure*}
	\centering
	\includegraphics[width=2.0\columnwidth]{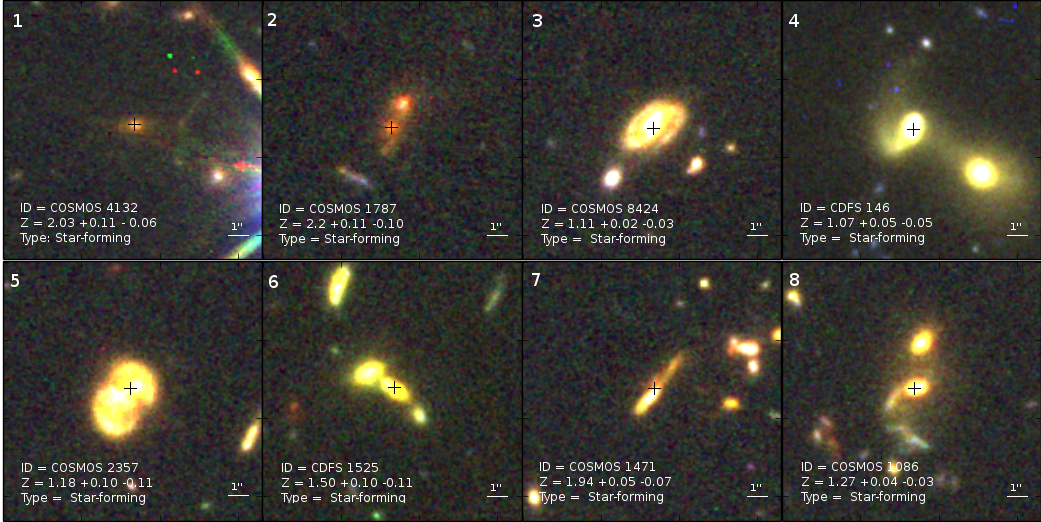}
	\caption{RGB thumbnails of composite sources using F125W, F160W and F814W filters, 3 sources show sure signs of merger activity (\# 4,5 \& 8), and 7/8 show nearby companions or distorted morphologies (with the exception of composite \#1)}
	\label{fig:Composites}
\end{figure*}

Of our high-luminosity radio-AGN, 20 are found to be embedded within star-forming hosts (``composites" hereafter). These sources contain large amounts of dust and would be easy to misclassify as quiescent without the use of our medium band near-infrared filters ({\it J1}, {\it J2}, {\it J3}, {\it Hs} and {\it Hl}). Comparing composites to the star-forming population we see in Figure \ref{fig:HybMass} that a small number show extremely high SFR and SSFRs and are undoubtedly undergoing a star-burst phase. Despite this, Figure \ref{fig:AGNProps} shows that overall the SFRs and dust content of composite sources remain consistent with the mass-similar star-forming population (P = 0.13 and 0.77 respectively). A visual inspection of Figure \ref{fig:Composites} shows that of the 8 composites covered by HST (CANDELS, \citealt{Grogin2011}), 7 have nearby companions and 3 show clear tidal interactions (objects 4, 5 and 8). However, these 3 sources do not correspond with any of the star-bursting objects visible in Figure \ref{fig:HybMass}: Bottom. Additionally, 4 composites are found to have strong X-ray counterparts with hardness ratios indicative of efficient accretion onto an AGN (Cowley et al, in prep) but also show no signs of elevated SSFRs. Finally, the high-luminosity composite population visible in ZFOURGE shows disk-like morphologies with a median Sersic index of 1.76 $\pm$ 0.18 \citep{VanderWel2014}. \\

Studying the evolution of high-luminosity radio-AGN as a whole, we can see that in terms of UVJ color, radio-AGN hosts evolve with redshift in a manner that is indistinguishable from their mass-similar, non-radio-AGN counterparts (Figure \ref{fig:AGNUVJ}). \\

\begin{figure*}
	\centering
	\includegraphics[width=2.0\columnwidth]{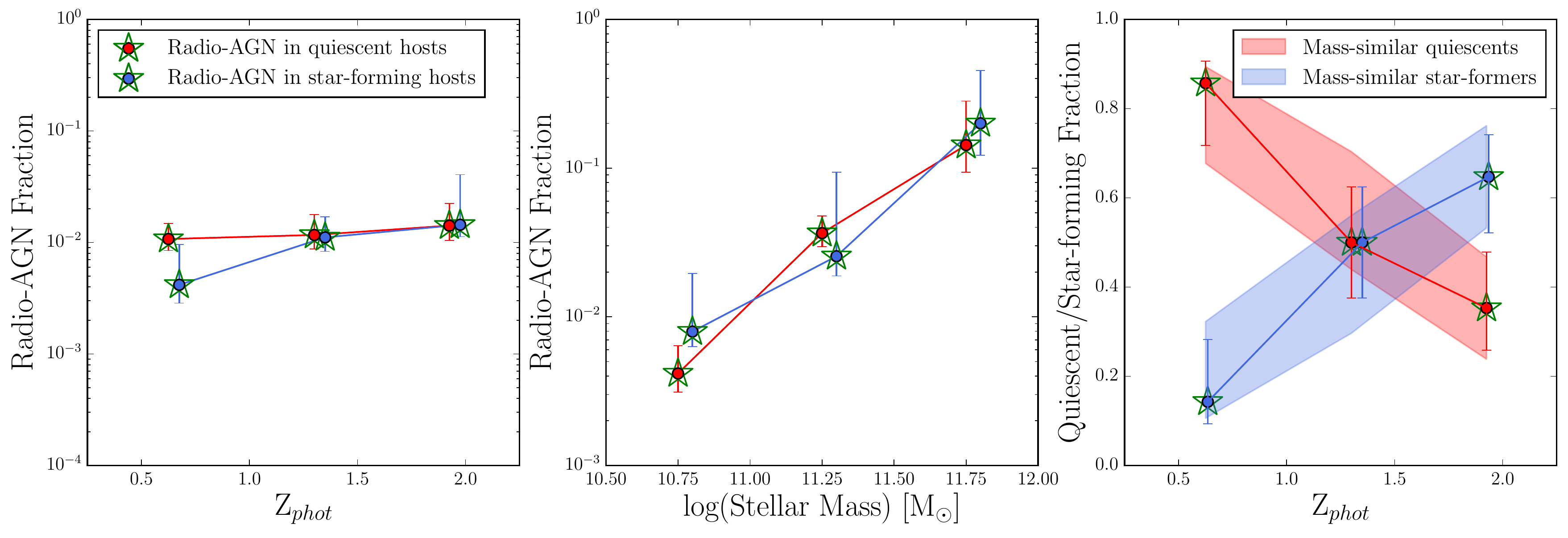}
	\caption{Left: The evolving radio-AGN fraction for quiescent (red with green stars) and star-forming (blue with green stars) high-luminosity radio-AGN above our mass limit of 10$^{10.5}$ M$_{\odot}$. Center: The radio-AGN fraction for each host type as a function of stellar mass shows a sharp drop-off in radio-AGN fraction as we progress to lower masses in excellent agreement with \citet{Best2005}. Left and Center panels show the number of quiescent or star-forming Radio-AGN hosts divided by the number of quiescent of star-forming galaxies in the given bin. Right: The breakdown of quiescent and star-forming high-luminosity radio-AGN hosts as a function of redshift. The fraction of quiescent (red shaded) and star-forming (blue shaded) galaxies of similar masses to the radio-AGN hosts is shown for comparison. All Errors are BETA confidence intervals and NMAD errors from 1000 random samples for the radio-AGN and mass-similar samples respectively.}
	\label{fig:AGNFrac}
\end{figure*}

Figure \ref{fig:AGNFrac}: Left, shows the fraction of galaxies containing a high-luminosity radio-AGN as a function of redshift. We find that across our full redshift range, an average of 1.0$^{+0.3}_{-0.2}$\% of quiescent, 1.1$^{+0.3}_{-0.2}$\% of star-former and only 1.0$^{+0.3}_{-0.1}$\% of all massive galaxies (M $\ge$ 10$^{10.5}$ M$_{\odot}$) host a high-luminosity radio-AGN, with little evolution in these values as a function of redshift. In Figure \ref{fig:AGNFrac} (Center), we see a tight dependence between high-luminosity radio-AGN and stellar mass, with a significant decrease in the abundance these objects from approximately 10\% at M$_{\odot}$ $\ge$ 10$^{11.5}$ down to less than 1\% at M$_{\odot}$ $\ge$ 10$^{10.5}$. We draw attention to the sharp drop-off seen in Figure \ref{fig:AGNFrac} (Right) where the fraction of star-forming high-luminosity radio-AGN hosts declines rapidly below a redshift of z = 1.5. This is consistent with the declining percentage of high-mass galaxies that are star-forming in general, within our mass-limited sample. This may explain the rarity of low redshift composites as simply due to the lack of suitable high-mass star-forming galaxies in the local universe. Finally, we note that in both Figure \ref{fig:AGNFrac}: Left and Right, the star-forming and quiescent high-luminosity radio-AGN fractions follow the same trend within errors. 

\section{Discussion} 
	\label{Sec:Discuss}

Using our high luminosity (L$_{1.4}$ $>$ 10$^{24}$ W~Hz$^{-1}$) mass-limited radio-AGN sample, we are able to study the fraction of massive galaxies that host radio-AGN to a redshift of z = 2.25. Figure \ref{fig:MassLim} shows that high luminosity radio-AGN become rare beneath a stellar-mass limit of $\sim$ 10$^{10.5}$ M$_{\odot}$ at low redshifts (z = 0.25). Interestingly this limit evolves with time, such that it mimics the expected evolution of the stellar-to-halo mass relation for a fixed halo-mass of 10$^{12}$ M$_{\odot}$. The implication of this result is that there may be a link between galaxies with high-mass halos and the triggering of high-luminosity radio-AGN. Indeed, the apparent halo-mass limit for hosting radio-AGN is consistent with the critical halo-mass at which simulations typically have to invoke AGN feedback in order to reproduce the local galaxy mass function \citep{Croton2006,Springel2005}. 

As seen in Figure \ref{fig:MassLim}, only a small number of high-luminosity radio-AGN fall below our 10$^{12}$ M$_{\odot}$ halo-mass line (converted to stellar mass using the \citet{Moster2010} stellar-to-halo mass relation. At z $>$ 1 this may be partially due to the incompleteness of NMBS below 10$^{10.5}$ M$_{\odot}$. However even at z $<$ 1, where our completeness is well above 80\% we see very few radio-AGN in low mass galaxies. Assuming that this halo-mass is required for powerful radio-AGN activity, the few objects that are found in lower mass galaxies may reflect scatter in the stellar-to-halo mass relation ($\simeq$ 0.2$\pm$ dex at z = 0 \citealt{More2009, Yang2009, Reddick2013, Behroozi2013}). In addition to this, \citet{Kawinwanichakij2014} find that quiescent galaxies at high redshift have unusually large halo-masses for a given stellar-mass (by up to $\simeq 0.1-0.2$ dex), hence it is possible that some of our low-mass radio-AGN hosts may actually reside in halos larger than the Moster et al median. \\

\begin{figure}
	\centering
	\includegraphics[width=1.0\columnwidth]{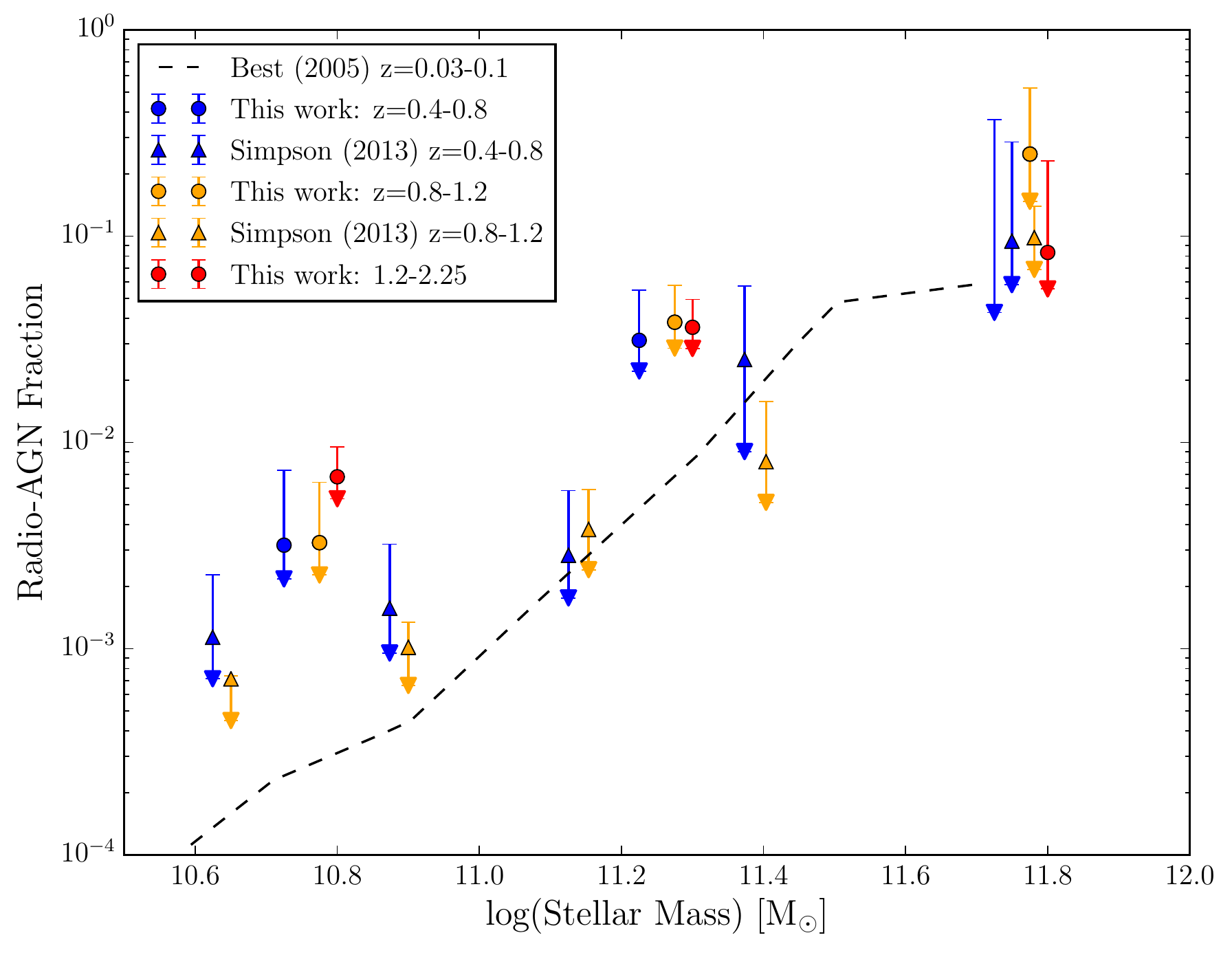}
	\caption{The fraction of galaxies containing radio-AGN with luminosities greater than 10$^{24}$ W~Hz as a function of mass for three redshift bins: 0.4 $<$ z $<$ 0.8 (blue circles), 0.8 $<$ z $<$ 1.2 (orange circles) and 1.2 $<$ z $<$ 2.25 (red circles). We see no evolution between our redshift bins. This is in good agreement with previous studies by Best et al (2005) at 0.03 $<$ z $<$ 0.1 (blue dashes) and Simpson et al (2013) at 0.4 $<$ z $<$ 0.8 and 0.8 $<$ z $<$ 1.2 (color coded as before). Our low-mass radio-AGN fraction is elevated at all redshifts compared to previous studies, which may be due to differences in how we select our radio-AGN samples.}
	\label{fig:Best}
\end{figure}

Alternatively, the observed halo-mass limit we discussed above may simply correspond to the stellar mass where the high-luminosity radio-AGN fraction becomes negligible. Figure \ref{fig:AGNFrac} (Left and Center) shows the fraction of massive galaxies containing a high-luminosity radio-AGN as a function of redshift and mass. While we find little evolution in the high-luminosity radio-AGN fraction as a function of redshift, we do observe a strong correlation with stellar-mass. Figure \ref{fig:Best} shows a comparison between the mass dependant radio-AGN fraction of our study to those of previous work in three redshift bins: 0.4 $<$ z $<$ 0.8, 0.8 $<$ z $<$ 1.2 (to match previous studies) and 1.2 $<$ z $<$ 2.25. Our high-luminosity radio-AGN fraction is slightly higher than that found in previous studies (0.03 $<$ z $<$ 0.1, \citealt{Best2005} and 0.4 $<$ z $<$ 1.2, \citealt{Simpson2013}) and we speculate that this may be due to differences in the techniques used to identify our radio-AGN samples. Despite this offset, the lack of evolution between redshift bins in our study is consistent with the lack of evolution seen between redshifts in these previous investigations. This raises the question of whether the universal radio-AGN fraction is simply a function of stellar-mass, with the most massive galaxies being much more likely to host radio-AGN activity than low-mass galaxies. In extremity, extrapolating upward in stellar-mass suggests that above stellar masses of 10$^{12}$ M$_{\odot}$, nearly all galaxies should contain a radio-AGN. \\

To investigate the impact of radio-AGN on host galaxy properties, we also use our mass-similar sample to compare radio-AGN hosts to non-hosts. We find that the high number of close companions and mergers seen in our high-luminosity composite (radio-AGN in star-forming hosts) sample seems to support a merger driven AGN triggering model. Despite this, we see no evidence for the enhanced star-formation rates expected during merger scenarios (Table 4) as composite SFRs are indistinguishable from the mass-similar star-forming sample. This may only reflect the short time-scales of a merger-induced star-burst. Finally, Figure \ref{fig:AGNFrac} (Right) shows that the fraction of radio-AGN hosted within quiescent and star-forming hosts evolves in a similar fashion to the non-AGN galaxy population. These results suggest that the radio-AGN and non-radio-AGN populations are hosted by galaxies with similar properties.

\section{Conclusions}
\label{Sec:Conclusion}

We have combined a deep {\it Ks}-band selected sample, which is ideal for selecting galaxies across a wide redshift range \citep{Rocca-Volmerange2013a}  with high quality photometric redshifts and sensitive 1.4~GHz VLA observations to identify 412 radio sources in the CDF-S and COSMOS legacy fields. This sample is split into two sub-groups: radio-AGN and radio detected star-forming galaxies using the ratio of the UV+24$\mu$m-based and radio-based star-formation rates. Using this sample we study the host properties of radio-AGN and compare them to a non-radio sample of similar masses and redshifts out to redshifts of 2.25. 

\begin{enumerate}

\item We find that the fraction of galaxies that contain radio-AGN at a given stellar mass shows little dependence on redshift or star-formation activity. This is in good agreement with the findings of \citet{Best2005} and \citet{Simpson2013} and extends the analysis of radio-AGN fractions to higher redshift ranges than previously explored. 
\\
\item The type of galaxies that host radio-AGN shows strong evolution as a function of redshift, from predominantly dusty, star-forming hosts in interacting or merger environments (1 $<$ z $<$ 2.25), to predominantly quiescent hosts at z $<$ 1. \\
\item This evolution is in line with the overall evolution of massive galaxies onto the red-sequence. \\
\item The above findings are in good general agreement with earlier work, particularly that of \citet{Rocca-Volmerange2013a} and \citet{DeBreuck2002}.
\\
\item Radio-AGN become particularly rare at halo masses below 10$^{12}$~M$_{\odot}$, suggesting that radio-AGN activity may be closely linked to high-halo masses. 
\\
\item Finally, our radio-AGN hosts show star-formation and specific star-formation rates consistent with non-radio-AGN hosts of similar mass and redshift, in good agreement with earlier work out to z $\sim$ 0.7 \cite{Johnston2008, Chen2013}.

\end{enumerate}

In summary we find that radio-AGN hosts show no statistical differences from non-hosts across a broad redshift range (0.25 $\le$ z $<$ 2.25) and that the fraction of galaxies containing radio-AGN at a given stellar mass is independent of both redshift and host type. Ultimately, by including these observations into models of AGN feedback and triggering, we hope that new insights may be gained into how these powerful objects both form and evolve.  \\

\label{lastpage}
\vspace{10mm}
We would like to thank the Mitchell family for their continuing support of the ZFOURGE project. We would also like to thank the Carnegie Observatories and the Las Campanas Observatory for providing the facilities and support needed to make ZFOURGE what it is today. Australian access to the Magellan Telescopes was supported through the National Collaborative Research Infrastructure Strategy of the Australian Federal Government. This research has made use of NASA's Astrophysics Data System. This research has made use of the NASA/IPAC Extragalactic Database (NED) which is operated by the Jet Propulsion Laboratory, California Institute of Technology, under contract with the National Aeronautics and Space Administration. This research made use of APLpy, an open-source plotting package for Python hosted at http://aplpy.github.com. This research made use of Astropy, a community-developed core Python package for Astronomy \citep{Robitaille2013}. This work made use of the IPython package \citep{PER-GRA:2007}. This research made use of matplotlib, a Python library for publication quality graphics \citep{Hunter:2007}. PyRAF is a product of the Space Telescope Science Institute, which is operated by AURA for NASA. This research made use of SciPy \citep{jones_scipy_2001}. GGK was supported by an Australian Research Council Future Fellowship FT140100933."

\bibliography{library}{}
\bibliographystyle{mn2e}
\appendix

\begin{landscape}
\begin{table}
\centering
\label{Tab:DataZFOURGE}
\begin{tabular}{@{}ccccccccccccccc@{}}
& & & & & & & & & & & \\
& & & & & & & & & & & \\
& & & & & & & & & & & \\
& & & & & & & & & & & \\
\hline
ID & RA & DEC & {\it Ks}-Band & z$_{phot}$ & z$_{spec}$ & Mass & F$_{1.4}$  & L$_{1.4}$  & L$_{IR}$ & L$_{UV}$ & SFR$_{Radio}$ & SFR$_{UV+IR}$  & SFR-Ratio & Type \\
   & [deg] & [deg] & [MAG] &  &  & [M$_{\odot}$] & [$\mu$Jy] & [W~Hz] & [W~Hz] & [W~Hz] & [M$_{\odot}$ yr$^{-1}$] & [M$_{\odot}$ yr$^{-1}$] & & \\
\hline 
COSMOS-10055 & 150.14721680 & 2.33718370 & 19.87 & 0.74 & 0.73 & 10.69 & 67.0 & 1.13e+23 & 9.96e+11 & 1.24e+10 & 3.58e+01 & 1.13e+02 & 0.32 & SF \\
COSMOS-10472 & 150.07937620 & 2.34056950 & 22.78 & 0.22 & - & 8.21 & 61.0 & 7.58e+21 & 1.42e+10 & 3.76e+07 & 2.41e+00 & 1.57e+00 & 1.54 & SF \\
COSMOS-1055 & 150.05722050 & 2.20592760 & 18.62 & 0.09 & 0.19 & 9.43 & 446.0 & 9.08e+21 & 4.09e+10 & 2.28e+08 & 2.89e+00 & 4.54e+00 & 0.64 & QU \\
COSMOS-1086 & 150.09982300 & 2.20327520 & 21.71 & 1.27 & - & 10.75 & 84.0 & 4.44e+23 & 1.27e+11 & 8.16e+09 & 1.41e+02 & 1.68e+01 & 8.42 & SF \\
COSMOS-1096 & 150.05662540 & 2.20855470 & 18.79 & 0.21 & - & 10.13 & 342.0 & 4.02e+22 & 5.69e+11 & 1.20e+09 & 1.28e+01 & 6.25e+01 & 0.20 & SF \\
COSMOS-11061 & 150.18797300 & 2.35292740 & 19.50 & 0.27 & - & 10.07 & 90.0 & 1.73e+22 & 7.78e+10 & 8.21e+08 & 5.50e+00 & 8.78e+00 & 0.63 & SF \\
COSMOS-11391 & 150.17570500 & 2.35869840 & 18.36 & 0.21 & - & 10.51 & 94.0 & 1.10e+22 & 1.23e+11 & 4.08e+09 & 3.50e+00 & 1.49e+01 & 0.23 & SF \\
COSMOS-11542 & 150.08515930 & 2.35732890 & 20.97 & 1.21 & - & 10.99 & 211.0 & 1.02e+24 & 5.51e+10 & 6.04e+09 & 3.24e+02 & 8.18e+00 & 39.61 & QU \\
COSMOS-11559 & 150.14303590 & 2.35588170 & 24.20 & 3.00 & - & 10.45 & 517.0 & 1.51e+25 & 1.38e+13 & 2.74e+10 & 4.81e+03 & 1.52e+03 & 3.18 & SF \\
\hline
\end{tabular}
\caption{Extract of the included data set. ID is the Field \& ID of the ZFOURGE {\it Ks}-band object, RA is the right ascension of the ZFOURGE {\it Ks}-band object in decimal degrees, DEC is the declination of the ZFOURGE {\it Ks}-band object in decimal degrees, z$_{phot}$ is the peak of photometric redshift probability distribution determined by EAZY for the ZFOURGE {\it Ks}-band object, Mass is the logged stellar mass of the ZFOURGE {\it Ks}-band object in M$_{\odot}$, F$_{1.4}$ is the 1.4~GHz flux of the ZFOURGE {\it Ks}-band object in micro-Janskys assuming association of sources within 1", L$_{1.4}$ is the 1.4~GHz luminosity of the ZFOURGE {\it Ks}-band object assuming association of sources within 1$\arcsec$ and a spectral index of 0.5 in WHz, L$_{IR}$ is the bolometric IR luminosity of the ZFOURGE {\it Ks}-band object assuming a Wuyts 2011 average SED template to extrapolate from 24um fluxes, L$_{UV}$ is the EAZY interpolated rest-frame 2800 $\AA$ luminosity of the ZFOURGE {\it Ks}-band object, SFR$_{Radio}$ is the 1.4~GHz luminosity of the radio object in M$_{\odot}$ per year, SFR$_{UV+IR}$ is the combined SFR from L$_{UV}$ \& L$_{IR}$ in M$_{\odot}$ per year and Type is the stellar population type of the {\it Ks}-band object based on our rest-frame UVJ color-color classification into Quiescent (QU) or Star-forming (SF) sources. The full table is available online.}
\end{table}

\begin{table}
\centering
\label{Tab:DataNMBS}
\begin{tabular}{@{}ccccccccccccccc@{}}
\hline
ID & RA & DEC & {\it Ks}-Band & z$_{phot}$ & z$_{spec}$ & Mass & F$_{1.4}$ & L$_{1.4}$ & L$_{IR}$ & L$_{UV}$ & SFR$_{Radio}$ & SFR$_{UV+IR}$ & SFR-Ratio & Type \\
   & [deg] & [deg] & [MAG] & & & [M$_{\odot}$] & [$\mu$Jy] & [W~Hz] & [W~Hz] & [W~Hz] & [M$_{\odot}$ yr$^{-1}$] & [M$_{\odot}$ yr$^{-1}$] & & \\

\hline 
NMBS-30222 & 149.75816471 & 2.18125926 & 19.66 & 0.93 & - & 10.76 & 83.0 & 2.60e+23 & 1.02e+11 & 6.16e+09 & 1.44e+02 & 1.30e+02 & 1.11 & SF \\
NMBS-33599 & 150.17541090 & 2.42608575 & 18.29 & 0.31 & 0.31 & 9.91 & 112.0 & 3.00e+22 & 2.03e+10 & 8.52e+09 & 1.66e+01 & 3.58e+01 & 0.46 & SF \\
NMBS-16854 & 150.00712980 & 2.45347805 & 18.65 & 0.76 & - & 11.23 & 761.0 & 1.54e+24 & 6.36e+08 & 3.75e+09 & 8.48e+02 & 6.59e+00 & 128.80 & QU \\ 
NMBS-16644 & 149.85016665 & 2.45223286 & 19.41 & 0.72 & 0.71 & 10.55 & 1102.0 & 1.98e+24 & 1.17e+10 & 5.65e+09 & 1.10e+03 & 4.78e+01 & 22.91 & SF \\
NMBS-13387 & 150.09534701 & 2.38475016 & 17.48 & 0.28 & 0.27 & 10.80 & 240.0 & 5.30e+22 & 9.06e+10 & 2.20e+09 & 2.93e+01 & 1.97e+01 & 1.49 & SF \\
NMBS-31704 & 149.86023730 & 2.29759213 & 20.77 & 1.83 & - & 11.12 & 68.0 & 9.41e+23 & 5.07e+10 & 2.39e+10 & 5.20e+02 & 3.77e+02 & 1.38 & SF \\
NMBS-11075 & 149.96143463 & 2.34943035 & 19.14 & 0.93 & - & 11.18 & 230.0 & 7.28e+23 & 2.07e+10 & 5.55e+09 & 4.02e+02 & 4.25e+00 & 94.59 & QU \\
NMBS-2925 & 149.74311397 & 2.21380815 & 19.04 & 0.87 & 0.89 & 11.21 & 309.0 & 8.41e+23 & 2.21e+10 & 4.11e+09 & 4.64e+02 & 4.91e+00 & 94.65 & QU \\
NMBS-6355 & 149.85677987 & 2.27315493 & 18.73 & 0.76 & 0.76 & 11.12 & 88.0 & 1.78e+23 & 2.21e+10 & 1.20e+10 & 9.81e+01 & 6.46e+01 & 1.52 & SF \\
NMBS-7479 & 149.88340095 & 2.29052069  &18.16 & 0.49 & 0.48 & 10.96 & 86.0 & 6.45e+22 & 2.13e+09 & 1.76e+09 & 3.56e+01 & 4.79e+00 & 7.43 & QU \\
NMBS-23410 & 150.05479754 & 2.56948127 & 19.97 & 0.76 & - & 10.37 & 162.0 & 3.24e+23 & 2.31e+10 & 5.78e+09 & 1.79e+02 & 1.56e+02 & 1.15 & SF \\ 
\end{tabular}
\caption{Extract of the included NMBS data set. As above. Full table is available online}
\end{table}

\end{landscape}

\begin{landscape}
\begin{table}
\centering
\label{Tab:DataMissed}
\begin{tabular}{@{}cccccccccccccccccccccccccc@{}}
& & & & & & & & & & & \\
& & & & & & & & & & & \\
& & & & & & & & & & & \\
& & & & & & & & & & & \\
\hline
RA [deg] & DEC [deg] & PF$_{1.4}$ [$\mu$Jy] & PE$_{1.4}$ [$\mu$Jy] & IF$_{1.4}$ [$\mu$Jy] & IE$_{1.4}$ [$\mu$Jy] & RMS [$\mu$Jy] & KsLimit [mag] & Sep [arcsec] & KsNearest [mag] & Cmnt \\
\hline 
149.7393292 & 2.4506194 & 69.0 & 12.0 & 69.0 & 12.0 & 12.0 & 23.73 & 3.49 & 23.05 & - \\
149.7405542 & 2.6053000 & 125.0 & 14.0 & 125.0 & 14.0 & 14.0 & 23.55 & 0.20 & 16.81 & - \\
149.7515542 & 2.5240667 & 46.0 & 13.0 & 46.0 & 13.0 & 13.0 & 24.02 & 4.44 & 23.65 & - \\
149.7698083 & 2.2180667 & 44.0 & 12.0 & 44.0 & 12.0 & 12.0 & 23.97 & 1.15 & 22.06 & - \\
149.7769625 & 2.4667833 & 50.0 & 13.0 & 50.0 & 13.0 & 13.0 & 24.19 & 2.00 & 21.43 & - \\
149.7876500 & 2.2293944 & 47.0 & 11.0 & 47.0 & 11.0 & 11.0 & 24.00 & 3.58 & 22.53 & - \\
149.7935083 & 2.2583306 & 57.0 & 10.0 & 57.0 & 10.0 & 10.0 & 23.81 & 2.48 & 22.78 & - \\
149.8471583 & 2.2152889 & 65.0 & 11.0 & 65.0 & 11.0 & 11.0 & 24.19 & 1.34 & 22.11 & - \\
149.8623292 & 2.4679250 & 41.0 & 11.0 & 41.0 & 11.0 & 11.0 & 24.19 & 4.66 & 21.34 & - \\
149.8673417 & 2.2416222 & 87.0 & 11.0 & 87.0 & 11.0 & 11.0 & 24.21 & 4.46 & 20.7 & - \\
149.8717583 & 2.2121917 & 67.0 & 12.0 & 67.0 & 12.0 & 12.0 & 24.20 & 4.07 & 21.84 & - \\
\hline
\end{tabular}
\caption{Extract of the included data set detailing radio sources that have no detected counterpart within the NMBS COSMOS observations. RA is the right ascension of the radio source in decimal degrees, DEC is the declination of the radio source in decimal degrees. PF$_{1.4}$ is the 1.4 GHz peak radio flux in micro-Janskys, PE$_{1.4}$ is the error on the 1.4 GHz peak radio flux in micro-Janskys, IF$_{1.4}$ is the integrated 1.4~GHz flux in micro-Janskys and IE$_{1.4}$ is the error on the integrated flux in micro-Janskys. RMS is the local RMS at the source in micro-Janskys, KsLimit is the limiting Ks magnitude at the source position in magnitudes with a zero-point of 25. Sep is the separation between the radio source and the nearest Ks detected object in arc-seconds and KsNearest is the magnitude of this nearest Ks detected object, in magnitudes with a zero-point of 25. Comnt is an index referring to any comments on the source which are located at the top of the catalogue. This data is taken from the COSMOS Deep project catalogues. The full tables for each of our fields {\it Ks} non-detections are available online.}
\end{table}

\end{landscape}

\end{document}